\documentclass[aps,prb,showpacs,twocolumn]{revtex4-1}

\usepackage[dvips]{graphicx}
\usepackage{amsmath,amssymb}
\usepackage{amsfonts}
\usepackage{color}

\usepackage[usenames,dvipsnames]{xcolor}

\usepackage[normalem]{ulem}   

\begin{document}

\newcommand{\Eq}[1]{Eq.~\eqref{#1}}
\newcommand{\Eqs}[1]{Eqs.~\eqref{#1}}
\newcommand{\obsWW}[1]{{\color{orange} \it [#1]}}
\newcommand{\modWW}[1]{{\color{red} #1}}
\newcommand{\VS}[1]{{\color{purple} #1}}
\newcommand{\ui}{\text{i}}
\newcommand{\im}{\text{Im}}
\newcommand{\re}{\text{Re}}
\newcommand{\iexp}{\text{i}}
\newcommand{\Tr}{\text{Tr}}
\newcommand{\la}{\langle}
\newcommand{\ra}{\rangle}
\newcommand{\uTHz}{\,\text{THz}}
\newcommand{\uGHz}{\,\text{GHz}}
\newcommand{\uMHz}{\,\text{MHz}}
\newcommand{\uW}{\,\text{W}}
\newcommand{\ueV}{\,\text{eV}}
\newcommand{\uF}{\,\text{F}}
\newcommand{\uH}{\,\text{H}}
\newcommand{\um}{\text{m}}
\newcommand{\uOhm}{\,\Omega}
\newcommand{\flq}{\left(\frac{\hbar}{2 e}\right)^2}
\newcommand{\fext}{f_{\text{ext}}}
\newcommand{\ELcav}{E_{L,\text{cav}}}
\newcommand{\df}{\delta\!f}
\newcommand{\Lcav}{L_{\text{cav}}}
\newcommand{\Ccav}{C_{\text{cav}}}
\newcommand{\Hcav}{H_{\text{cav}}}

\newcommand{\knd}{k_n d}
\newcommand{\kpd}{k_0 d}
\newcommand{\kd}{k d}
\newcommand{\wndelta}{(\omega_n+\delta)}

\newcommand{\dk}{\delta_k}
\newcommand{\Gd}{G(\delta_k)}
\newcommand{\bw}{\text{b}}

\newcommand{\xn}{\left(k_n \frac{d}{2} + \beta_n\right)}
\newcommand{\yn}{\left(k_n \frac{d}{2} - \beta_n\right)}

\newcommand{\Gmean}{\sqrt{\Gamma_1\Gamma_2}}

\newcommand{\deltath}{\delta_{\rm th}}

\title{Non-degenerate parametric resonance in tunable superconducting cavity}
\author{Waltraut Wustmann$^{1,2}$}
\author{Vitaly Shumeiko$^{1}$}
\affiliation{$^1$Chalmers University of Technology, S-41296 G\"oteborg, Sweden\\
$^2$Laboratory for Physical Sciences, College Park, MD 20740, USA}
\date{10 April 2017}

\begin{abstract}

We develop a theory for non-degenerate parametric resonance in a tunable superconducting cavity. We focus on nonlinear effects that are caused by nonlinear Josephson elements connected to the cavity. We analyze parametric amplification in a strong nonlinear regime at the parametric instability threshold, and calculate maximum gain values. Above the threshold, in the parametric oscillator regime the linear cavity response diverges at the oscillator frequency at all pump strengths. We show that this divergence is related to the continuous degeneracy of the free oscillator state with respect to the phase.  Applying on-resonance input lifts the degeneracy and removes the divergence. We also investigate the quantum noise squeezing. It is shown that in the strong amplification regime the noise undergoes four-mode squeezing, and that in this regime the output signal to noise ratio can significantly exceed the input value. We also analyze the intermode frequency conversion and identify parameters at which  full conversion is achieved.

\end{abstract}

\maketitle

\section{Introduction}
Quantum parametric resonance in superconducting Josephson circuits finds numerous applications  in circuit-QED   
technology. A novel generation of quantum limited parametric amplifiers
\cite{CasLeh2007,CasETAL2008,YamETAL2008,MartinisAPL2013,BergealNature2010,SiddiqiPRB2011,VionPRB2014,SimoenJPL2015,SiddiqiScience2015} 
makes possible  single shot readout and continuous monitoring of states of superconducting qubits\cite{SiddiqiPRL2011,SiddiqiNature2013a,NakamuraAPL2013,DiCarloNature2013,KranzNatComm2016}.   Noise squeezing under parametric down-conversion \cite{EichlerPRL2011} is used to enhance the qubit coherence time\cite{SiddiqiNature2013,SiddiqiPRX2016}.  Among other applications are  efficient generation of entangled microwave photons\cite{BergealPRL2012,FluETAL2012,RocETAL2012,EichlerPRL2012,MenzMarx2012} and intermode frequency conversion \cite{AumentadoNaPh2011,AbdoPRL2013}. 

A detailed theory of degenerate parametric resonance in a tunable superconducting cavity was developed in Ref. \cite{WustmannPRB2013}. 
In this paper we extend this theory to the regime of non-degenerate resonance, when two cavity modes with frequencies $\omega_n$ and $\omega_m$ are coupled by parametric pumping with frequency $\Omega \approx \omega_n\pm\omega_m$. A tunable superconducting cavity is a resonator integrated with a superconducting quantum interferometer device (SQUID)  that serves as a variable inductance controlled by magnetic flux. \cite{Sandberg2008,Walquist2006}. Variation of the SQUID inductance changes the cavity resonance frequencies, and the parametric resonance is excited by rapid modulation of the SQUID inductance with appropriate frequency (parametric flux pumping). 

As it was discussed in Ref.~\cite{WustmannPRB2013} a high gain amplification regime in the vicinity of the parametric instability threshold is strongly nonlinear. This nonlinearity limits the gain and squeezing at the threshold, and moreover  it 
saturates the parametric instability and establishes a stationary oscillator regime above the threshold \cite{WilsonNature2011}. In contrast to optical parametric amplifiers and oscillators where nonlinearity is typically related to pump depletion \cite{GrahamZPh1968,MilburnBook}, the nonlinearity of the tunable cavity is introduced by the nonlinear inductance of the SQUID. In a small amplitude limit this is a cubic, Duffing type nonlinearity of the Josephson current-phase dependence, which plays the role of the Kerr effect for the cavity field.

The non-degenerate parametric amplification possesses interesting new features compared to the degenerate case.
Amplification of weak input follows a well known two-mode squeezing scenario of the linear parametric amplification theory   \cite{CavesPRD1982,ScullyBook,ClerkRMP2010}. However, the cavity linear response under the presence of strong on-resonance intracavity field generated, e.g. by amplified coherent signal or self-sustained parametric oscillation,  exhibits a four-mode squeezing. This effect is explained by the  intracavity field acting as a secondary, ``current'' pump that excites additional idlers via four-mode mixing. As we will show, the signal to noise ratio in this regime can be significantly enhanced compared to the input. 

Furthermore, the cavity linear response in the parametric oscillator regime  diverges at the oscillation frequency at all pump strengths. This phenomenon is analogous to the one in the optical parametric oscillators, which  attracted great deal of attentions \cite{GrahamZPh1968,FabreJdP1989,BjorkPRA1988,DrummondPRA1989}  (for more recent discussions see e.g. Ref.~\cite{NussenzveigOC2004,ValcarcelPRA2010}).
This divergence is closely related to the continuous degeneracy of the free oscillator state with respect to the oscillation phase, and it is lifted by applying an on-resonance input. 
  
The structure of the paper is the following. In Section \ref{TunableCavity} we briefly outline the description of the tunable cavity developed in paper \cite{WustmannPRB2013}, and introduce dynamical equations for the non-degenerate parametric resonance. Next Section~\ref{Amplification} is devoted to the parametric amplification regime. First we consider the parametric instability and parametric oscillation. Then we proceed with a classical theory of nonlinear amplification across the parametric threshold. Sections~\ref{2modelinear} and E are devoted to the detailed study of the  linear response of the empty cavity, relevant for the noise squeezing, and  the multimode response of the filled cavity using the framework of supermodes \cite{BraunsteinPRA2005,TrepsD2010}. Here we discuss the effect of continuous degeneracy of parametric oscillation and phase locking by means of weak signal injection. 
In Section \ref{Fluctuation} we study the  quantum noise squeezing; here we compute quadrature correlation functions and analyze the signal to noise ratio for linear and nonlinear amplification regimes. In the end of the section we present explicit equations for squeezed vacuum below parametric threshold. In Section \ref{Conversion}  we derive the scattering matrix for the parametric frequency conversion.

\section{Tunable cavity}
\label{TunableCavity}

\begin{figure}[h]
\centering
 \includegraphics[width=0.9\columnwidth]{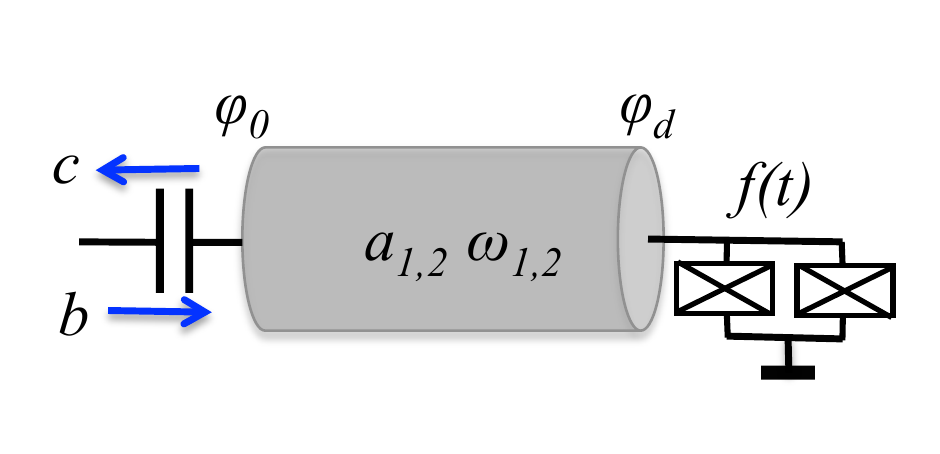}
\caption{Sketch of tunable cavity. $a_n(t)$ is the complex amplitude of $n$-th cavity eigen mode, $b(t)$ and $c(t)$ are input and output field amplitudes, respectively; the cavity eigen frequencies $\omega_n$  are controlled by magnetic flux $f(t) = F + \df  \cos\Omega t$. }\label{Fig:TunableCavity}
\end{figure}

The tunable cavity we study in this paper is a $\lambda/4$ superconducting resonator connected at one end to a SQUID, and at the other end to a transmission line, see Fig.~\ref{Fig:TunableCavity}. The parametric effect  is achieved by rapid temporal modulation of a magnetic flux through the SQUID, which  results in the variation of the boundary condition at the cavity edge that shifts the cavity resonance frequencies. 

In this paper, we consider the non-degenerate parametric resonance, which corresponds to the modulation with frequency  close to the sum, or difference, of the cavity eigenmode frequencies, $\Omega \approx  \omega_2 \pm \omega_1$. Our present study is built on the detailed analysis of this device in the context of the degenerate parametric regime~\cite{WustmannPRB2013}. Here we briefly outline the results of that analysis.

The quantum Hamiltonian of the tunable cavity derived in Ref.~\cite{WustmannPRB2013} has the form,
\begin{eqnarray}
\label{eq:Hcav}
&& \Hcav(a_n) = \sum_n \hbar \omega_n a_n^\dag a_n  + V(a_n,t) \,, \\
&& V(a_n,t) =
- (E_J \sin F)( \df  \cos\Omega t) \, \phi_d^2 - \frac{E_J \cos F}{12}  \phi_d^4 \,.  \nonumber
\end{eqnarray}
This Hamiltonian describes the field inside the cavity, $\phi(x,t)$, in terms of the cavity eigen modes, 
\begin{eqnarray}\label{eq:modeexpansion}
&&\phi (x,t) = \sqrt{(2e)^2\over \hbar \Ccav }
\sum_{n=1}^\infty  { \cos(k_n x)\over \sqrt{\omega_n }} (a_n(t) + a_n^\dag(t))
\,.
\end{eqnarray}
 In these equations, $E_J$ is the Josephson energy of the SQUID junction, $F$ is the constant magnetic flux bias, 
$\df$ is the amplitude of the flux temporal modulation;  $\phi_d(t)$ refers to the field boundary value at the cavity edge, $x=d$, connected to the SQUID; $\Ccav$ is the cavity capacitance, and $k_n = \omega_n/v$ is the mode eigenvector,
$v=d/\sqrt{C_{cav}L_{cav}}$ is the electromagnetic wave velocity; 
$a_n(t)$ are the  mode annihilation operators,  which satisfy the bosonic commutation relations, 
$[a_n\,,\, a_m^\dag\,] = \delta_{nm}$.
 
The cavity mode spectrum is defined by the equation,
\begin{eqnarray}
\label{eq:dispersion}
(\knd)\tan \knd  =  \frac{2 E_J \cos F}{\ELcav} \, = {1\over \gamma}  
\,,
\end{eqnarray}
where  $\gamma$ is a participation ratio of the inductive energies of the cavity, 
$\ELcav = (\hbar/2e)^2(1/\Lcav)$, and the SQUID. The participation ratio is small, $\gamma \ll 1$, for magnetic flux bias not too close to $\pi/2$. A weak effect of the Josephson junction capacitance, which is small compared to the cavity capacitance $\Ccav$, is neglected here.
The non-equidistant character of the cavity spectrum allows the selective parametric excitation of only two cavity modes. 

Important constraints under which the Hamiltonian (\ref{eq:Hcav}) is valid concern small values of the amplitudes of the parametric modulation and the field at the cavity edge, 
\begin{eqnarray}\label{constraints}
\df,\; \phi_d  \ll 1 \,.
\end{eqnarray}
%

\subsection{Equation of motion}

The quantum dynamics of the field in the cavity is described by the set of Langevin equations associated with the Hamiltonian (\ref{eq:Hcav}),
\begin{eqnarray}\label{eq:Langa}
i \dot a_n - \omega_n a_n - {1\over \hbar} [a_n \,,\,V(a_n,t)]   + i\Gamma_n a_n
\nonumber\\
= \sqrt{2\Gamma_{n0}} \,b(t)
\,.
\end{eqnarray}
These equations take into account  external losses related to the coupling to the  transmission line\cite{WustmannPRB2013},
\begin{equation}\label{eq:Gn0}
 \Gamma_{n0} = \omega_n \left( \frac{C_c}{\Ccav} \right)^2 k_n d \,, 
\end{equation}
where $C_c$ is the coupling capacitance, $\Gamma_n$ refers to the total  losses.
The operator $b(t)$ indicates an input field defined through the mode operators of the transmission line, $a_k(t)$ \cite{ColGar1984,YurkeBook}, 
\begin{eqnarray}
b(t) = \sqrt{v\over 2\pi } \int _{0}^{\infty} dk \,a_k(t_0)e^{-i\omega_k(t-t_0)} \,.
\end{eqnarray}

The output field operator $c_n(t)$ is related to the input operator via the relation \cite{ColGar1984},
\begin{equation}\label{eq:cb}
c_n(t) = b(t) - i\sqrt{2\Gamma_{n0}} \,a_n(t) 
 \;.
\end{equation}
%

\subsection{Resonance approximation}

The non-degenerate parametric resonance has two qualitatively different regimes. The first, amplification or down-conversion regime corresponds to the pump frequency close to the sum of the frequencies of the two selected modes,
\begin{eqnarray}\label{Omegadown}
\Omega = \omega_2 + \omega_1 + 2\delta, \quad \delta \ll \omega_n \,.
\end{eqnarray}
For the amplification regime a natural reference frame is a rotating frame with frequencies $\omega_{1,2} + \delta$. 
In this frame the mode operators undergo the transformations,
\begin{eqnarray}
a_n \rightarrow  e^{-\iexp(\omega_n+\delta) t}  a_n\,.
\end{eqnarray}
The corresponding transformation of the cavity Hamiltonian
is produced by a unitary operator, 
\begin{eqnarray}
&&U= \exp[-\iexp (\omega_1+\delta)a_1^\dag a_1 - \iexp (\omega_2+\delta)a_2^\dag a_2] \\
&& \Hcav \to \Hcav - \ui \hbar \dot U U^\dag = \Hcav - \hbar \sum_{n} \wndelta a_n^\dag a_n.
\nonumber
\end{eqnarray}
Averaging over rapid oscillation results in the reduced Hamiltonian of the resonance approximation, 
\begin{eqnarray}\label{eq:Heff_A}
&& \Hcav = -\sum_{n=1,2} \left[ \hbar \delta  a_n^\dag a_n 
 +  {\hbar \alpha_n\over2} \left(a_n^\dag a_n + {1\over 2}\right)^2 \right] \\
 &&  - 2\hbar \alpha \left(a_1^\dag a_1 + \frac{1}{2}\right) \left(a_2^\dag a_2 + {1\over2}\right) 
 - \hbar  \epsilon \left( a_1 a_2 +  a_1^\dag a_2^\dag \right) 
  \nonumber
 \,.
\end{eqnarray}
Here we have defined  mode-specific nonlinearity parameters,
\begin{eqnarray}
\label{eq:def_alpha_n}
 \alpha_n 
 &=& \frac{\hbar}{2\gamma \ELcav} 
 \left(\frac{\sqrt{\omega_n} \cos{k_n d}}{k_n d} \right)^4 \,,\\
  \alpha &=& \sqrt{\alpha_1 \alpha_2} \,,
\end{eqnarray}
and a pump strength
\begin{eqnarray}
\label{eq:def_epsilon}
  \epsilon 
 &=& \frac{ \df \tan{F}}{2\gamma} 
 \left(\frac{\sqrt{\omega_1} \cos k_1 d}{ k_1 d}\right)
 \left(\frac{\sqrt{\omega_2} \cos k_2 d}{ k_2 d}\right)
 \,.
\end{eqnarray}

The resonance approximation for the Langevin equations yields two coupled equations,
\begin{eqnarray}
\label{eq:EOM_A1_ampl}
 \ui \dot a_n  &+&  \delta a_n + \ui \Gamma_n a_n + \epsilon a_m^\dag   
 +  \alpha_n \left(a_n^\dag a_n + 1\right) a_n \nonumber \\
  &+& 2 \alpha \left(a_m^\dag a_m + \frac{1}{2}\right) a_n = \sqrt{ 2\Gamma_{n0}} \,b_n(t) \,,
\end{eqnarray}
where $m\neq n$, and the input fields, $b_{n}(t)$, are written in the respective rotating frames, $\omega_{n} + \delta$.
The input-output relations in \Eq{eq:cb} retain their form in the rotating frame.

The alternative regime of parametric frequency conversion or up-conversion  corresponds to the pump frequency close to the difference of the frequencies of two selected modes, 
\begin{eqnarray}\label{Omegaup}
\Omega = \omega_2 - \omega_1 + 2\delta, \quad 
\omega_2 > \omega_1
\,.
\end{eqnarray}
In this case no amplification occurs while the resonant modes are hybridized, and the energy is transferred from the one mode to the other.
For this regime a natural reference frame is a rotating frame with frequencies, 
$\omega_{1} - \delta$, $\omega_{2} + \delta$, i.e.  
\begin{eqnarray}
a_1 \rightarrow  e^{-\iexp(\omega_1-\delta) t} a_1, \quad 
a_2 \rightarrow  e^{-\iexp (\omega_2 + \delta) t} a_2
\,.
\end{eqnarray}
 The corresponding unitary operator is, 
\begin{eqnarray}
U = \exp[-\iexp (\omega_1-\delta) a_1^\dag a_1 - \iexp (\omega_2+\delta) a_2^\dag a_2] \,.
\end{eqnarray}
The resonant Langevin equations in the rotating frame take the form,
\begin{eqnarray}
\label{eq_upc:EOM_A1_hyb}
 i \dot a_n &\mp& 
 \delta a_n + \ui \Gamma_n a_n + \epsilon a_m + \alpha_n \left(a_n^\dag a_n + 1\right) a_n \nonumber \\
 &+& 2 \alpha \left(a_m^\dag a_m + \frac{1}{2}\right) a_n = \sqrt{ 2\Gamma_{n0}} \,b_n(t)\,,
\end{eqnarray}
where the upper and lower signs refer to mode 1 and mode 2, respectively.

The resonance approximation implies a slow resonance dynamics on the time scale set by the cavity eigen frequencies, 
\begin{eqnarray}\label{constraints2}
\delta, \;\epsilon, \; \alpha_n|A_n|^2, \; \Gamma_n \ll \omega_n
\end{eqnarray}
(here $A_n$ indicates a quasiclassical intracavity field). Bearing in mind that for small $\gamma$ the estimate, $\cos k_nd \sim \gamma\ll1$, holds for low frequency modes and bias flux $F$ not particularly close to $\pi/2$, and using \Eqs{eq:modeexpansion}, (\ref{eq:def_epsilon}), and (\ref{eq:def_alpha_n}), we are able to obtain relations, 
\begin{eqnarray}\label{constraints3}
{\epsilon\over\omega_n} \sim \gamma\, \df \tan F  \ll 1, \;  \; {\alpha_n|A_n|^2\over \omega_n}
\sim \gamma\phi_d^2 \ll 1\,.
\end{eqnarray}
These inequalities  respect constraints in \Eq{constraints2}, and they are automatically fulfilled by virtue of the constraints in \Eq{constraints}. 
Furthermore, relations \Eq{constraints2} provide a room for the validity of the theory well above the parametric oscillation threshold for a high quality cavity, 
\begin{eqnarray}
\Gamma_n &\ll& \epsilon \sim \omega_n\gamma\, \df \tan F \ll \omega_n\,, \nonumber\\
 \Gamma_n &\ll&  \alpha_n|A_n|^2 \sim  \omega_n \gamma\phi_d^2 \ll \omega_n\,.
\end{eqnarray}
%

\section{Parametric amplification}
\label{Amplification} 
 
We start our study with the classical amplification regime.
By denoting classical fields with capital letters, $A_n$, we write a classical version of \Eq{eq:EOM_A1_ampl} in the form,
\begin{eqnarray}
\label{eq:EOM_class_ampl}
 \ui \dot A_1  + (\zeta_1  + \ui \Gamma_1) A_1 + \epsilon A_2^\ast   &=& \sqrt{ 2\Gamma_{10}} B_1(t)\\
 - \ui \dot A_2^\ast + (\zeta_2 - \ui \Gamma_2 ) A_2^\ast + \epsilon A_1  &=&
   \sqrt{ 2\Gamma_{20}} B_2^\ast(t)  \nonumber 
\,,
\end{eqnarray}
where
\begin{eqnarray}\label{eq:zetas}
\zeta_1 &=& \delta +\alpha_1 |A_1|^2 + 2 \alpha |A_2|^2 \nonumber\\
\zeta_2 &=& \delta+ \alpha_2 |A_2|^2 + 2 \alpha |A_1|^2 
\,.
\end{eqnarray}
These terms describe the nonlinear  self-Kerr effect proportional to $\alpha_n$, and the cross-Kerr effect proportional to $\alpha$.

\subsection{Parametric instability and oscillation}
\label{Instability}

\begin{figure}[tbh]
\begin{center}
\includegraphics[width=\columnwidth]{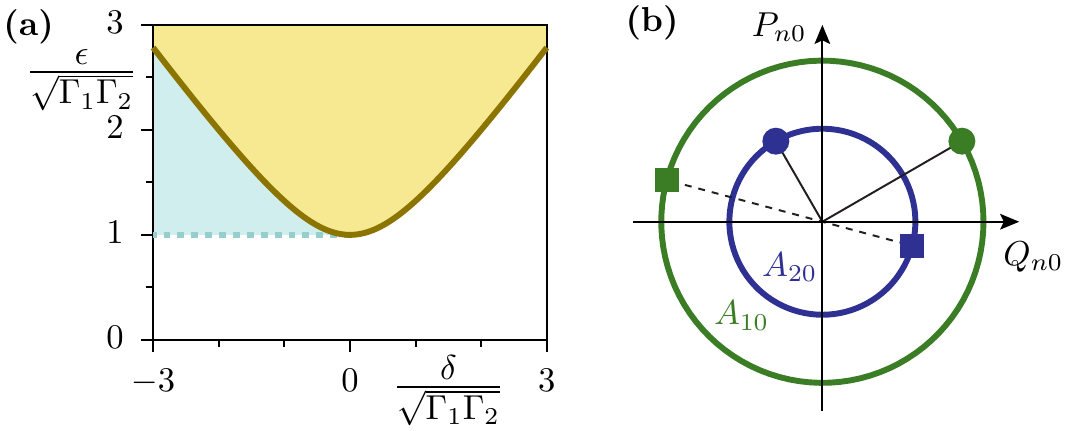}
\caption{
(a) Region of parametric oscillation in ($\epsilon-\delta$) plane (yellow), 
and coexistence region of parametric oscillation and empty cavity states (blue).
(b) Quadratures $Q_{n0} = (A_{n0} + A_{n0}^\ast)/2$ and $P_{n0} = (A_{n0} - A_{n0}^\ast)/(2\ui)$
of parametric oscillation state forming circles, with radii determined by \Eqs{eq:Asq1_over_Asq2__LC}--(\ref{A10}), and phases constrained by \Eq{eq:Theta}; indicated are the state pairs $(Q_n,P_n)$ for two specific values of phase difference, $\psi = -\pi/2$ (circles) and $\psi = \pi$ (squares), the phase sum $\Theta = 5\pi/6$ remains invariant.
($\epsilon=2\Gmean$, $\delta=0$,
$\Gamma_2=3\Gamma_1$, $\alpha_2=3\alpha_1=3\sqrt{\Gamma_1\Gamma_2}/100$)
}
\label{fig:radiation_quadratures}
\end{center}
\end{figure}

First we consider the dynamics of the closed cavity. The empty cavity state, $A_n=0$, always exists but looses stability  within a certain region of the pump strengths and detunings.  To perform the stability analysis we evaluate the eigen frequency spectrum of the linearized equation (\ref{eq:EOM_class_ampl}). 
Assuming $\zeta_n = \delta$, and $A_1(t), \,A_2^\ast(t) \propto e^{-i\Delta t}$,   we compute the determinant of the dynamical matrix, 
\begin{eqnarray}\label{Det=0}
{\rm Det} =  (\Delta + \delta+ i\Gamma_1)(-\Delta + \delta- i\Gamma_2) -\epsilon^2 =0 \,.
\end{eqnarray}
Within the stability region the solution to this equation must have a negative imaginary part, ${\rm Im}\,\Delta \leq 0$, and therefore the condition, ${\rm Im} \,\Delta = 0$, must define the boundary of this region, i.e.~the instability threshold. 
Using this argument  we solve \Eq{Det=0} separately  for the real and imaginary parts and find  the threshold,
\begin{eqnarray}\label{eq:epsilon_thresh}
 \epsilon^2 & = & \Gamma_1 \Gamma_2 + \delta^2 \left[ 1 - \left(\frac{\Gamma_1 -\Gamma_2 }{\Gamma_1+\Gamma_2}\right)^2 \right] 
 \\ \label{eq:Delta_thresh}
 \Delta &=& {\Gamma_1  - \Gamma_2  \over \Gamma_1 + \Gamma_2}\, \delta \,.
\end{eqnarray}
As one sees from these equations, the instability occurs at frequencies generally deviating from  the cavity resonances. The instability threshold has a minimum value at the zero pump detuning $\delta$, where it is defined by the damping, $\epsilon=\sqrt{\Gamma_1\Gamma_2}$; the threshold grows with the pump detuning, as illustrated in Fig.~\ref{fig:radiation_quadratures}(a). 

In terms of the pump detuning, the  cavity  ground state is unstable within the interval,
\begin{eqnarray}\label{delta_thresh}
\delta^2 <  \deltath^2 = \frac{(\Gamma_1+\Gamma_2)^2}{4 \Gamma_1 \Gamma_2}\,
 (\epsilon^2-\Gamma_1 \Gamma_2)
\;.
\end{eqnarray}
For modes with equal dampings, $\Gamma_1 = \Gamma_2$, the threshold value becomes identical to the one of the degenerate  case \cite{WustmannPRB2013}, and the deviation of the critical fluctuation frequency turns to zero, $\Delta=0$.

The instability leads to the emergence of a self-sustained parametric oscillation above the threshold, 
and it is described with a nonlinear solution of homogeneous \Eq{eq:EOM_class_ampl}.  This regime 
is manifested by spontaneous radiation from the cavity at two frequencies shifted from the cavity 
resonances, $\omega_1+\delta+\Delta_0$ and $\omega_2+\delta-\Delta_0$. This is different from  the 
degenerate case, where the parametric radiation frequency coincides with half of the pump frequency, $\Omega/2$, 
and coincides with the cavity mode frequency at zero pump detuning, $\delta = 0$. 

The oscillation frequency 
shift is derived in Appendix~\ref{sec:limitcycle_downconversion}, 
\begin{eqnarray}\label{eq:Delta0}
 \Delta_{0} = {\Gamma_1 \zeta_2 - \Gamma_2 \zeta_1 \over \Gamma_1 + \Gamma_2} 
\;.
\end{eqnarray}
It grows with increasing pump intensity, and vanishes only if the modes have identical characteristics, $\Gamma_1=\Gamma_2$ and $\alpha_1=\alpha_2$. Following terminology of Ref.~\cite{FabreJdP1989} we will call such modes balanced.

The oscillation is characterized by complex amplitudes, $A_{n0}= |A_{n0}|e^{i\theta_n}$, whose moduli are related,
\begin{equation}\label{eq:Asq1_over_Asq2__LC}
 \frac{| A_{20}|^2}{ |A_{10}|^2} = \frac{\Gamma_1}{\Gamma_2}
 \,,
\end{equation}
as it is found in Appendix~\ref{sec:limitcycle_downconversion}, and  
\begin{eqnarray}\label{A10}
|A_{10}|^2 = {2(-\delta \mp \delta_{\rm th}) \Gamma_2\over \alpha_1\Gamma_2 + \alpha_2\Gamma_1
+ 2\alpha(\Gamma_1 + \Gamma_2)}\,.
\end{eqnarray}
Similar to the degenerate case, \Eq{A10} describes  unstable (upper sign)  and stable (lower sign) oscillator states. 
The stable state exists at all  $\delta<\delta_{\rm th}$, and it coexists with the stable trivial state at $\delta<-\delta_{\rm th}$ (see Fig.~\ref{fig:radiation_quadratures}a); in the latter region both the excited and trivial stable states are separated by an unstable state.
For the balanced modes  \Eq{A10} reduces to the one for a degenerate oscillator (Eq.~(45) in Ref.~\cite{WustmannPRB2013}),
\begin{eqnarray}\label{A0}
|A_{0}|^2 = {-\delta \mp \sqrt{\epsilon^2-\Gamma^2}\over 3\alpha}\,,
\end{eqnarray}
with rescaling, $\alpha \to 3\alpha$, that stems from the cross-Kerr effect.  

The properties of the oscillation phases $\theta_n$  are  qualitatively different from the degenerate case: there the phase takes two values differing by $\pi$, implying double degeneracy of the oscillator state. Here the oscillator state has a continuous degeneracy with respect to the difference, $\psi=\theta_1-\theta_2$, of the mode phases. 
The sum of the phases is fixed and for the stable state reads, according to \Eqs{sin}, (\ref{cos}),
\begin{eqnarray}\label{eq:Theta}
&&\sin\Theta= {\sqrt{\Gamma_1\Gamma_2}\over \epsilon}, \quad 
\cos\Theta = - {\sqrt{\epsilon^2-\Gamma_1\Gamma_2}\over \epsilon}, \\
&& \Theta = \theta_1 + \theta_2  \in (\pi/2, \pi) \,. \nonumber
\end{eqnarray}

The output radiation is connected to the intracavity field via relation, 
\begin{eqnarray}
C_{n0} = |C_{n0}| e^{i\theta_{Cn}} = - i\sqrt{2\Gamma_{n0}} \,A_{n0}, 
\end{eqnarray}
then the output radiation phases are related,
\begin{eqnarray}
\theta_{C1} + \theta_{C2} = \Theta -\pi \,,
 \end{eqnarray}
as well as the radiation intensities,
\begin{equation}\label{eq:Csq1_over_Csq2__LC}
 \frac{| C_{20}|^2}{ |C_{10}|^2} = \frac{\Gamma_1\Gamma_{20}}{\Gamma_2\Gamma_{10}}
 \;.
\end{equation}
In the ideal cavity, $\Gamma_n=\Gamma_{n0}$, the radiation intensities are equal in both modes.

\subsection{Nonlinear gains }

Now we switch on an input in \Eq{eq:EOM_class_ampl}, and suppose a harmonic  input in the first mode, $B_1(t) = B_1(\Delta) e^{-i\Delta t}$, slightly detuned, by $\Delta$, from the reference frame, 
$\omega_1+\delta$, and having a complex amplitude, $B_1(\Delta)$. Below the parametric threshold, this signal would generate an intracavity field consisting of two harmonics, 
 the signal, $A_1(t) = A_1 (\Delta)e^{-i\Delta t}$, and the idler,  $A_2(t) = A_2 (-\Delta)e^{i\Delta t}$. 
The idler is detuned by $-\Delta$ from its reference frequency, $\omega_2+\delta$.
The same harmonic components will be present in the output field, $C_1(t) = C_1(\Delta)e^{-i\Delta t}$ and 
$C_2(t) = C_2(-\Delta)e^{i\Delta t}$. 

This arrangement may also include a second input signal at the idler frequency, $B_2(t) = B_2(-\Delta)e^{i\Delta t}$. Then \Eq{eq:EOM_class_ampl} reduces to a static equation
for intracavity field amplitudes,
\begin{eqnarray}\label{AB}
(\Delta + \zeta_1 +   i \Gamma_1) A_1(\Delta) + \epsilon A_2^\ast (-\Delta)&=& \sqrt{2\Gamma_{10}} B_1(\Delta) \nonumber\\
(- \Delta  +\zeta_2  - i \Gamma_2) A_2^\ast (-\Delta)+ \epsilon A_1(\Delta) &=& \sqrt{2\Gamma_{20}} B_2^\ast(-\Delta) 
\,.\nonumber\\
\end{eqnarray}
Inverting these equations yields explicitly the intracavity field,
\begin{eqnarray}\label{eq:Aampl} 
&& \left(
\begin{array}{c}
  A_1 (\Delta)     \\
  A_2^\ast   (-\Delta)  
\end{array}
\right) 
= {\cal A}(\Delta)
\left(
\begin{array}{c}
\sqrt{2\Gamma_{10}}  \,B_1(\Delta)      \\
 \sqrt{2\Gamma_{20}}  \, B_2^\ast (-\Delta)    
\end{array}
\right)
\,,
\end{eqnarray}
where
\begin{eqnarray}\label{A}
&& {\cal A}(\Delta) = {1\over {\rm Det}}
 \left(
\begin{array}{cc}
-\Delta  + \zeta_2 - i\Gamma_2  & -\epsilon   \nonumber\\  
-\epsilon  & \Delta+ \zeta_1  + i\Gamma_1     
\end{array}
\right)\\
\label{eq:det}
&& {\rm Det}(\Delta) = (\Delta+ \zeta_1 +i\Gamma_1)(-\Delta + \zeta_2 -i\Gamma_2) - \epsilon^2
\,.
\end{eqnarray}
This formal solution for the intracavity field together with \Eq{eq:cb} allows us to formulate the input-output relation, the Bogoliubov transformation,
\begin{equation}\label{C}
\left(
\begin{array}{c}
  C_1 (\Delta)     \\
  C_2^\ast(-\Delta)
\end{array}
\right)
= {\cal V}(\Delta)%
\left(
\begin{array}{c}
  B_1(\Delta)      \\
  B_2^\ast   (-\Delta)  
\end{array}
\right)
\,.
\end{equation}
The input-output matrix elements are,
\begin{eqnarray}\label{V}
{\cal V}_{11}(\Delta) &=& 1 - {2i \Gamma_{10}  ( -\Delta-i\Gamma_2 +\zeta_2) \over {\rm Det}(\Delta) } \nonumber \\
\label{eq:Vmatrixelements}
{\cal V}_{22} (\Delta)&=& 1 + {2i  \Gamma_{20}  ( \Delta+i\Gamma_1 +\zeta_1) \over {\rm Det}(\Delta) } \\
{\cal V}_{12} (\Delta)&=&   {2 i \epsilon\sqrt{\Gamma_{10}\Gamma_{20}}\over {\rm Det}(\Delta) }  = -{\cal V}_{21}(\Delta) 
\,.
\nonumber
\end{eqnarray}
One can straightforwardly check that matrix ${\cal V}(\Delta)$ possesses the properties, 
\begin{equation}\label{eq:Vproperty}
|{\cal V}_{11}|^2 - |{\cal V}_{12}|^2 = 1, \quad |{\cal  V}_{22}| = |{\cal  V}_{11}|, \quad |{\cal  V}_{12}| 
= |{\cal  V}_{21}|
\,, 
\end{equation}
in the absence of internal losses, $\Gamma_n= \Gamma_{n0}$.

Amplification of a single mode input, $B_1(\Delta)$, is characterized by the gains,
\begin{eqnarray}\label{eq:gains}
G_{11}(\Delta) = \left|{C_1(\Delta)\over B_1(\Delta)}\right|^2 = |{\cal V}_{11}(\Delta)|^2, \nonumber\\
G_{12}(-\Delta) = \left|{C_2(-\Delta)\over B_1(\Delta)}\right|^2 =  |{\cal V}_{21}(\Delta)|^2.
\end{eqnarray}
These gains are {\it nonlinear} functions of the input due to dependence on the intracavity field entering the Kerr terms, 
$\zeta_{n}(A_1, A_2)$ in \Eq{V}. Nevertheless, the nonlinear gains respect the relations,
\begin{eqnarray}\label{Gains_relations}
G_{11}(\Delta) &=& 1 + G_{12}(-\Delta)  \\
G_{11}(\Delta) &=& G_{22}(-\Delta), \quad G_{21}(\Delta) = G_{12}(-\Delta) \,, \nonumber
\end{eqnarray}
following from \Eq{eq:Vproperty}, which are the same  as the ones known from theory of linear amplifiers
\cite{CavesPRD1982,ClerkRMP2010}.

As we will see later in this section, the cavity response becomes increasingly nonlinear while 
approaching the instability threshold, where  amplification of even a single photon input becomes strongly 
nonlinear.

The application of the nonlinear \Eq{AB} above the parametric threshold requires certain care. In this region the stationary response is only possible for an input whose frequency coincides with the frequency of parametric oscillation, $\Delta = \Delta_0$. For detuned inputs the nonlinear response is non-stationary due to the mixing of the input and oscillation fields by the Kerr effect. If  the input is so weak that its contribution to the nonlinear terms can be neglected, the linearized response is stationary.

\subsection{Amplification of on-resonance signal}
In this section we analyze the nonlinear amplification of a single on-resonance input signal  to get
 a general picture of the nonlinear amplification during transition through the parametric oscillation 
 threshold. The results in this section are presented for relatively weak input power, $|B_1|^2 \lesssim \Gmean$, and neglecting intrinsic losses, $\Gamma_n = \Gamma_{n0}$. 
 

\subsubsection{Balanced modes}
\label{onresonance_balanced}

We first  investigate numerically a simpler case of  balanced modes,  
($\alpha_n=\alpha$, and $\Gamma_n=\Gamma$).  In this case the 
parametric oscillation has zero detuning,  
$\Delta_0=0$, and we explore amplification of the on-resonance signal with $\Delta=0$, both below and above the threshold.

\begin{figure}[tb]
\includegraphics[width=\columnwidth]{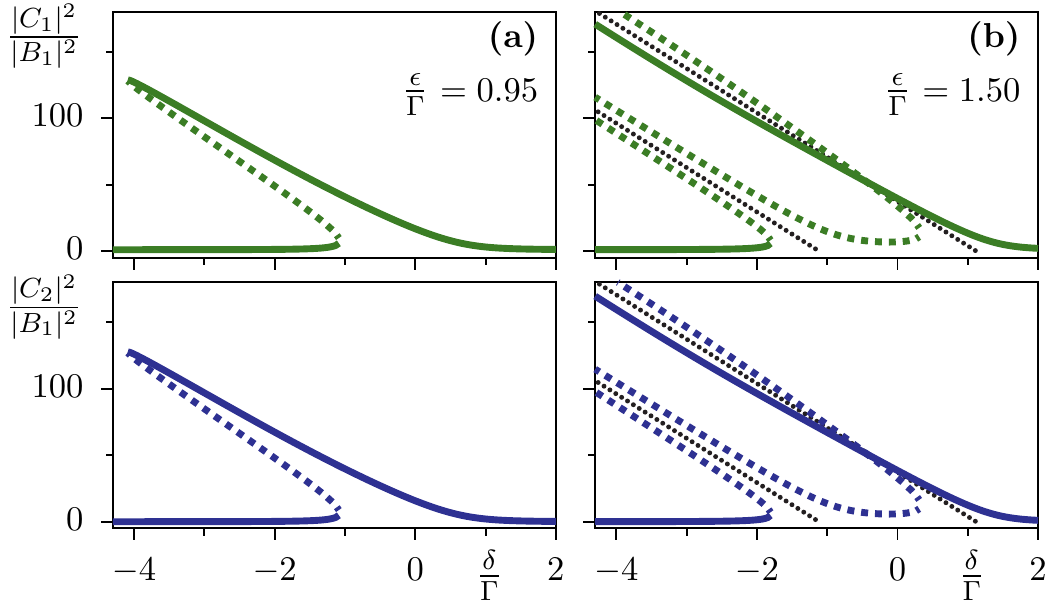}
\caption{
Nonlinear gains for signal and idler, $|C_{1,2}|^2$ (upper and lower panels, respectively) for undetuned, $\Delta=0$  input signal $B_1$,
following from \Eqs{eq:Aampl}, \eqref{eq:cb} for balanced modes, 
$\Gamma_1=\Gamma_2$, and $\alpha_1=\alpha_2$, as functions of  pump detuning $\delta$ for different values of the pump strength $\epsilon$: (a)
 below threshold $\epsilon/\Gamma=0.95$,  and (b)
 above threshold $\epsilon/\Gamma=1.5$. 
Solid (dashed) lines indicate (in)stable states; 
thin dotted lines indicate radiation amplitudes from stable and unstable parametric oscillator states. 
($|B_1|^2 = 2 \Gamma$, $\theta_B=0$, $\alpha=\Gamma/100$, $\Gamma = \Gamma_0$.)
}
\label{fig:nlin_response}
\end{figure}

In Fig.~\ref{fig:nlin_response} the computed cavity responses of both modes, $|C_{1,2}|^2$,
are presented as functions of the pump detuning $\delta$ below and above the parametric threshold.
The characteristic responses of both modes are qualitatively similar and resemble 
the response of the degenerate parametric amplifier (cf. Fig.~5 and discussion in Ref.~\cite{WustmannPRB2013}).

For $\epsilon$ smaller than the minimum parametric threshold, $\epsilon<\Gamma$,
the response curves $|C_{1,2}|^2(\delta)$ in Fig.~\ref{fig:nlin_response}(a) essentially describe a Duffing resonance. When approaching the threshold, $\epsilon \lesssim \Gamma$, the maxima of $|C_{1,2}|^2(\delta)$ are strongly enhanced, while the resonance width decreases owing to the reduced effective damping, $\Gamma^2 \to \Gamma^2 - \epsilon^2$.

The similarity to the degenerate resonance case is however illusive and does not reflect the fact that the intracavity dynamics here occurs in 
a higher-dimensional phase space and is more complicated. We attribute this similarity to the fact that 
the cavity steady state here remains close to the manifold characterized by $|A_{1}| = |A_{2}|$, which amounts to a projection to the degenerate subspace. This is a rather special situation which stems from the mode symmetry.

When $\epsilon > \Gamma$,  Fig.~\ref{fig:nlin_response}(b), the resonance splits in two branches, each branch consisting of a pair of steady states. 
The lower amplitude pair is formed close to the amplitude of the unstable parametric oscillation state
(dotted line) and both states of this pair are unstable as well. 
The higher amplitude pair is formed close to the amplitude of the stable parametric oscillation state (dotted line), but only one of its two components is stable. 
This is in constrast to the degenerate parametric amplifier where both components of the higher amplitude pair are stable, (Fig.~5(c-d) in Ref.~\cite{WustmannPRB2013}). 

The difference can be understood from the underlying parametric oscillation states, from which the branches emerge:
the degenerate parametric oscillator has two degenerate stable oscillation states which are $\pi$-shifted in phase. When a small external signal is applied each of these states remains stable being only shifted in the  quadrature plane. 
In contrast, the non-degenerate parametric oscillator has infinitely many stable oscillation states which are degenerate with respect to both $|A_1|^2$ and $|A_2|^2$, but differ by an arbitrary phase difference  $\theta_{1}-\theta_{2}$.
A small external signal breaks this rotational symmetry since $\theta_{1}-\theta_{2}$ acquires a fixed, $B$-dependent value according to \Eq{eq:Aampl} (see \Eq{A1A2symmetric} below, and further analytical details in Sec.~\ref{Phase locking}).

\subsubsection{Unbalanced modes}
In this section we consider the more realistic case of unbalanced modes with non-identical parameters. 
The mode parameters are specified by \Eqs{eq:Gn0} and (\ref{eq:def_alpha_n}), and in the limit, $\gamma \ll 1$, they have the scaling, 
\begin{eqnarray}\label{scalingG}
{\Gamma_{n0}\over \Gamma_{m0}}   =  {\alpha_n\over\alpha_m} \approx \left({2n-1\over 2m-1}\right)^2 .
\end{eqnarray}

To explore this more complex case we will have to take into account the nonzero detuning of the parametric oscillator frequency $\Delta_0$, and therefore consider input that is on-resonance with the oscillator, $\Delta = \Delta_0(\delta)$, above the threshold, while keeping for consistency the finite detuning,  $\Delta = \Delta_0(\delta_{th})$, below the threshold. 

\begin{figure}[tb]
\includegraphics[width=\columnwidth]{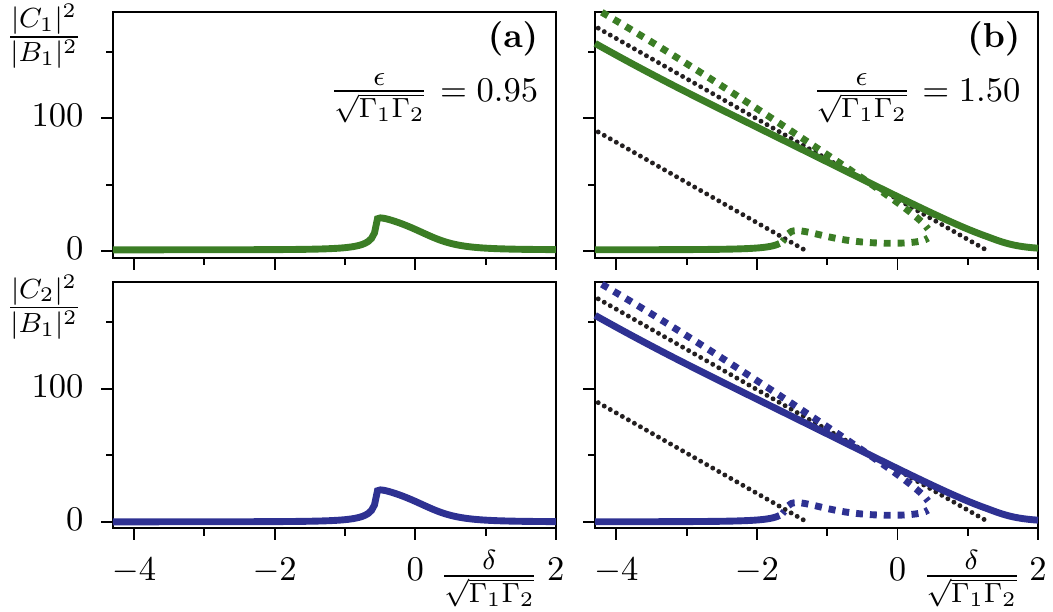}
\caption{
Nonlinear gains for signal and idler, $|C_{1,2}|^2$ (upper and lower panels, respectively) for unbalanced modes, $\Gamma_2=3\Gamma_1$, $\alpha_2=3\alpha_1$, as functions of  pump detuning $\delta$ for different values of the pump strength $\epsilon$: (a)
 below threshold $\epsilon/\sqrt{\Gamma_1\Gamma_2}=0.95$,  and (b)
 above threshold $\epsilon/\sqrt{\Gamma_1\Gamma_2}=1.5$. In (b) the input is detuned,  
 $\Delta = \Delta_0(\delta)$, to be on resonance with the parametric radiation.
 Thin dotted lines indicate radiation amplitudes from stable and unstable parametric oscillator states.
($|B_1|^2 = 2 \sqrt{\Gamma_1\Gamma_2}$, $\theta_B=0$,
$\alpha_1 = \sqrt{\Gamma_1\Gamma_2}/100$,
$\Gamma_{n} = \Gamma_{n0}$.)
}
\label{fig:nlin_response2}
\end{figure}

The most important implication of the mode asymmetry is that the internal cavity dynamics is no longer 
confined to the vicinity of the manifold $|A_{1}| = |A_{2}|$ in phase space. 
This has consequences for the cavity response already below threshold, as demonstrated in Fig.~\ref{fig:nlin_response2}(a).
Here the effectively reduced damping at strong pumping, 
$\epsilon^2 \sim \Gamma_1\Gamma_2$ 
in \Eq{eq:det}, does not result in a strongly enhanced Duffing resonance. Therefore the cavity does not 
automatically enter a regime of multistability, as it does in case of the degenerate parametric amplifier at small input.
This is related to the fact that the internal cavity amplitudes $|A_{1,2}|^2$ differ strongly although the illustrated output amplitudes $|C_{1,2}|^2$ are practically identical.

Above threshold, Fig.~\ref{fig:nlin_response2}(b), we still observe a stable high amplitude cavity state extending far at red detuning, $\delta < \deltath$,  but the unstable branch is confined to a small window of red-detuned $\delta$, Fig.~\ref{fig:nlin_response2}(b). 

%

\subsubsection{Nonlinear amplification close to threshold}\label{sec:nlin@threshold}
Some analytical results can be obtained for the regime of strong nonlinear amplification in close vicinity of the threshold. In this regime the width of the resonance is determined by the nonlinear frequency shifts, $\alpha_n |A_n|^2$, while these terms can still be relatively small,
\begin{eqnarray}\label{zeta<Gamma}
\left|1 - {\epsilon^2\over \Gamma_1\Gamma_2}\right|
\ll {\alpha_n |A_n|^2\over \Gamma_n} \ll 1 \,.
\end{eqnarray}

This regime starts below the threshold, and it persists across the threshold
as long as the contribution of the input to the Kerr terms dominates over the one of the emerging 
parametric oscillation. 
In principle this regime is relevant for any value of pump and input detunings, but we focus, 
for the sake of simplicity, on the case, $\delta=\Delta=0$, and neglect the internal losses, 
$\Gamma_n = \Gamma_{n0}$.

Then the solution for the intracavity field, \Eq{A}, takes an approximate form (for a single input
$B_1(0)$), 
\begin{eqnarray}\label{A@threshold}
&&A_1 = {-i\Gamma_2\over {\rm Det}}\sqrt{2\Gamma_1} B_1, \quad 
A_2^\ast = {-\epsilon\over {\rm Det}}\sqrt{2\Gamma_1} B_1  \nonumber\\
&&{\rm Det} = \zeta_1\zeta_2 + i(\Gamma_1\zeta_2 - \Gamma_2\zeta_1) \,.
\end{eqnarray}
From these equation we deduce,  
\begin{eqnarray}\label{A1toA2}
\left|{A_2\over A_1}\right|^2 = {\epsilon^2\over \Gamma_2^2} \approx {\Gamma_1\over \Gamma_2} \,.
\end{eqnarray}
Using this relation we compute the second term in the determinant in \Eq{A@threshold},
\begin{eqnarray}
i(\Gamma_1\zeta_2 - \Gamma_2\zeta_1)= i \,{|A_1|^2\over \Gamma_2}(\alpha_2\Gamma_1^2 - \alpha_1\Gamma_2^2) \,.
\end{eqnarray}
For unbalanced modes this term dominates over the product, $\zeta_1\zeta_2$, 
by virtue of \Eq{zeta<Gamma}, and we get the solution, using the scaling of \Eq{scalingG},
\begin{eqnarray}\label{eq:A1approx@threshold}
|A_1|^6 = {2\over (1-\Gamma_1/\Gamma_2)^2} \left({\Gamma_1\over\alpha_1}\right)^2 
{|B_1|^2\over \Gamma_1} \,.
\end{eqnarray}
Using this result, we are able to evaluate the magnitudes of the gains, \Eqs{eq:Vmatrixelements}--\eqref{eq:gains} at the threshold, $\epsilon^2 = \Gamma_1\Gamma_2$,
\begin{eqnarray}\label{Gnonlinear}
G_{11} \approx G_{12} = \left|{2\Gamma_1  \Gamma_2  
\over {\rm Det} }\right|^2 = 2|A_1|^2{\Gamma_1\over
|B_1|^2 } \,.
\end{eqnarray}

Combining equations,  \Eqs{zeta<Gamma} and (\ref{eq:A1approx@threshold}), we identify the conditions for the nonlinear regime to occur in terms of input power and pump strength,
\begin{eqnarray}\label{eqtmp:limits_nonlinearampl}
1 - {\epsilon^2\over \Gamma_1\Gamma_2} \ll \left[ {\alpha_1 \over \Gamma_1}{|B_1|^2 \over \Gamma_1}\right]^{1/3} \ll 1 . \;
\,
\end{eqnarray}
The window for the nonlinear regime is controlled by parameter $\Gamma_1/\alpha_1$.  The left inequality defines the nonlinear regime, and it is convenient to rewrite it as a lower bound on the input intensity,
\begin{eqnarray}\label{nonlinearity}
\left(1 - {\epsilon^2 \over \Gamma_1\Gamma_2} \right)^3 
{ \Gamma_1 \over \alpha_1}   \ll {|B_1|^2 \over \Gamma_1}  \,. 
\end{eqnarray}
The nonlinear amplification regime may start rather far from the threshold, for instance, for  a single photon input, $|B_1|^2/\Gamma_1=1$,  and $\Gamma_1/ \alpha_1 = 10$, it starts at $\epsilon > 0.77\sqrt{\Gamma_1\Gamma_2}$.  
The right constraint in \Eq{eqtmp:limits_nonlinearampl} allows sufficient room for the theory to be valid well above the single photon input, 
$1 \ll |B_1|^2/\Gamma_1 \ll \Gamma_1/ \alpha_1$, when $\Gamma_1/ \alpha_1 \gg 1$. 

The balanced mode case presents particular interest for the further discussion in the next sections. In this case, 
the intracavity fields have equal absolute values, $|A_1|=|A_2|=|A|$, according to \Eq{A1toA2}, 
which confirms our numerical observation in Fig.~\ref{fig:nlin_response}. 
The determinant, \Eq{A@threshold}, is given by the first term, 
${\rm Det} = (3\alpha|A|^2)^2$, since the second term disappears, and then
\begin{eqnarray}\label{A1A2symmetric}
A_1 &=& |A|e^{-i\pi/2+i\theta_B}, \quad  A_2 = |A|e^{i\pi-i\theta_B}\,, \nonumber\\
 |A|^5 &=& {\sqrt 2 \Gamma^{2}\over (3\alpha)^2} \,{|B_1|\over \sqrt{\Gamma}} \,.
\end{eqnarray}
The nonlinear gains for the balanced modes are determined by this equation together with \Eq{Gnonlinear}.
%

\subsection{Two-mode linear amplification}
\label{2modelinear}
Amplification of arbitrary detuned input signals can be analyzed in great detail, both analytically and numerically, in the linear  amplification regime. The results of this analysis are also relevant for evaluation of quantum noise and will be used later in Sec.~\ref{Fluctuation}.

In the linear regime the intracavity fields generated by the weak input are assumed to be small, $\alpha_n |A_n|^2 \ll \Gamma_n, \sqrt{\Gamma_1\Gamma_2- \epsilon^2}$, and the Kerr effect is neglected, $\zeta_n \approx \delta$. 
Then the linearized Bogoliubov transformation of \Eqs{C} and (\ref{V}),  can be written in the form, 
\begin{eqnarray}\label{CB}
&&  C_n(\Delta) =  u_n({\Delta}) B_n(\Delta) +  { v}_{n}({\Delta}) B^\dag_{m}(-\Delta)
\, 
\end{eqnarray}
($m \neq n$), where the coefficients, $u_1(\Delta) = {\cal V}_{11}(\Delta)$ and $v_1(\Delta) = {\cal V}_{12}(\Delta)$,
for the first mode have the explicit form,
\begin{eqnarray}\label{uv}
&&  u_{1}(\Delta) = { 
\left(\delta +\Delta +  i(\Gamma_1-2\Gamma_{10})\right)
\left(\delta -\Delta -  i\Gamma_2\right)    - \epsilon^2  \over 
\left(\delta +\Delta +  i\Gamma_1\right)
\left(\delta -\Delta -  i\Gamma_2\right)    - \epsilon^2 } \nonumber\\
&&  v_{1}(\Delta) = {  2i\sqrt{\Gamma_{10}\Gamma_{20}} \,\epsilon  \over 
\left(\delta +\Delta + i\Gamma_1\right)
\left(\delta -\Delta - i\Gamma_2\right)    - \epsilon^2} \,.
\end{eqnarray}
For the second mode the coefficients are obtained by permutation, $1 \leftrightarrow 2$. 

The coefficients satisfy the same relation as in \Eq{eq:Vproperty} 
\begin{eqnarray}\label{uvproperty}
&& |u_{n}(\Delta)|^2 - |v_{n}(\Delta)|^2 = 1 \nonumber\\
&&u_1(\Delta)v_2(-\Delta)- v_1(\Delta)u_2(-\Delta) = 0 \,.
\end{eqnarray}
These relations not only relate the signal and idler gains, 
$G_{11}(\Delta) - G_{12}(-\Delta) =1$, but also guarantee preservation of the bosonic commutation relations in the quantum regime \cite{CavesPRD1982}.

The linear gains diverge at the parametric instability threshold, 
 \Eqs{eq:epsilon_thresh}, \eqref{eq:Delta_thresh}, as is to be expected.

\begin{figure}[tbh]
\includegraphics[width=\columnwidth]{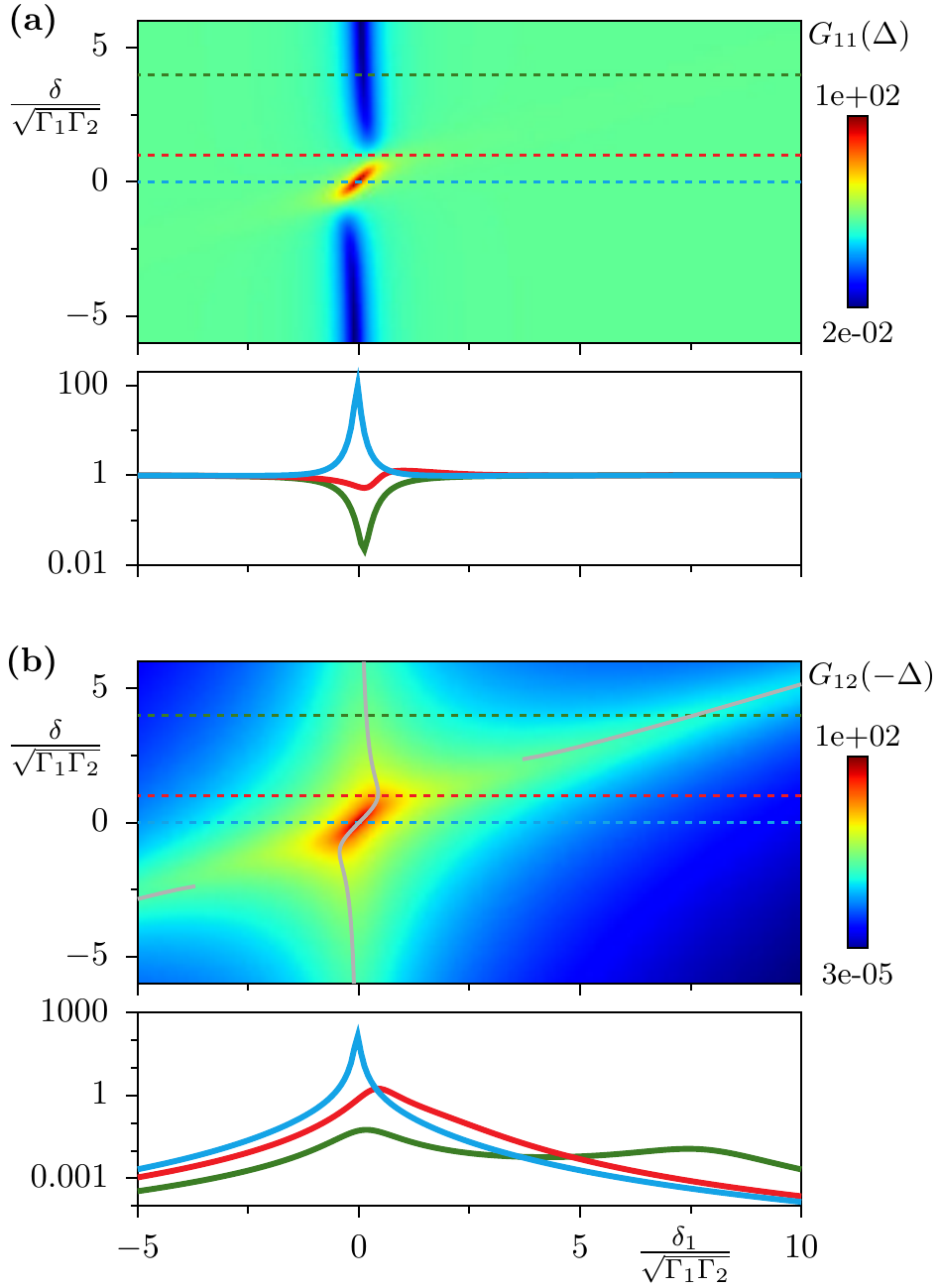}
\caption{
Linear gain spectra for signal and idler for detuned input below threshold. 
(a) Signal gain $G_{11}(\Delta)$,  and  (b) idler gain $G_{12}(-\Delta)$ as functions of pump detuning, $\delta$, and the signal detuning, $\delta_1 = \delta+\Delta$; 
horizontal dashed lines in color plots indicate cuts presented in lower panels, thin grey lines in (b) indicate positions of resonant peaks.
($\epsilon = 0.95\Gmean$, $\Gamma_{20}=3\Gamma_{10}$, 
$\Gamma_{1} = 1.8\Gamma_{10}$,
$\Gamma_{2} = 4\Gamma_{10} = (4/3) \Gamma_{20}$,
$\alpha_2=3\alpha_1$,
$\alpha_1 = \sqrt{\Gamma_{10}\Gamma_{20}}/100$.)}
\label{fig:lineargain_dw1_dpD_belowthresh}
\end{figure}
%

The linear gain spectra for signal, $G_{11}(\Delta) = |u_1(\Delta)|^2$, and idler, $G_{12}(-\Delta) = |v_2(-\Delta)|^2$, determined by \Eq{uv}, are illustrated in Fig.~\ref{fig:lineargain_dw1_dpD_belowthresh}. To facilitate a comparison with the experiment, we plot the spectra as functions of the input signal detuning 
$\delta_1$ from the cavity mode, $\omega_s = \omega_1 + \delta_1$ rather than the detuning  
$\Delta$ which is more convenient for analytics. Their relation is, $\delta_1 = \delta+ \Delta$. The idler gain is also plotted vs input signal detuning $\delta_1$, although the idler is detected at frequency, 
$\omega_i = \omega_2 + \delta - \Delta = \omega_2 + 2\delta - \delta_1$. Here we keep finite internal losses, $\Gamma_n \neq \Gamma_{n0}$.

For a pump far detuned from the resonance, $|\delta| > \Gmean$, the signal gain
 in Fig.~\ref{fig:lineargain_dw1_dpD_belowthresh}(a) shows a dip centered at $\delta_1=0$
due to the presence of internal losses  (green curve on the lower panel); the dip disappears when the signal is detuned away from the resonance, $\delta_1> \Gmean$, and the scattering is elastic, $G_{11}=1$. 

At zero pump detuning, $\delta=0$, the parametric amplification dominates over internal losses, giving rise to a strong gain peak, $G_{11}(\delta_1 \approx 0) \gg 1$ (blue curve).
In the intermediate region of $\delta$ there is a competition between the two effects. 
Since the amplification resonance occurs along a tilted
line in the ($\delta_1,\delta$)--plane, it shifts the internal loss resonance away from $\delta_1=0$
which therefore features an ``avoided crossing'' around $(\delta_1,\delta)=0$.
Within the avoided crossing both internal loss resonance 
and amplification resonance can coexist in the gain $G_{11}(\delta_1)$, at slightly shifted values of $\delta_1$  (red curve).

The parametric amplification resonance itself is better resolved in the 
gain spectrum of the idler, $G_{12}$,  Fig.~\ref{fig:lineargain_dw1_dpD_belowthresh}(b).
Since the idler frequency is out of resonance from the input at $\omega_1+\delta_1$,
it is not affected by the internal loss resonance,
and the $G_{12}$-spectrum is therefore characterized by the parametric amplification alone. 
The parametric amplification resonances  are determined by the local minima of the denominator in \Eq{uv}. 
For small pump detunings, $|\delta| < \Gmean$, a single resonance exists
approximately on the tilted line $\delta_1 = \delta + \delta(\Gamma_1-\Gamma_2)/(\Gamma_1+\Gamma_2)$ (blue curve on the lower panel)).
For larger values of $|\delta|$ this resonance approaches the line $\delta_1=0$ again
while at the same time the spectrum develops a shoulder into the blue-detuned (red-detuned) region for $\delta>0$ ($\delta<0$) (red curve), from which eventually a second resonance peak arises 
(green curve).
For balanced modes, $\Gamma_1=\Gamma_2=\Gamma$,
the resonances would be identical to those found for the degenerate parametric amplifier,\cite{WustmannPRB2013}
namely a single resonance at $\delta_1 = \delta$ [$\Delta=0$] for 
$|\delta| \leq \sqrt{\epsilon^2 + \Gamma^2}$,
and two split resonances, $\delta_1 = \delta \pm \sqrt{\delta^2 - \epsilon^2 - \Gamma^2}$, 
at $\delta >\sqrt{\epsilon^2 + \Gamma^2}$.

\subsection{Four-mode linear amplification}

Now we turn to the linear response in the presence of a strong intracavity field. Such a field can be generated either by an  input signal with another frequency or by parametric oscillation. In this case the amplification picture changes qualitatively. The strong microwave field in the cavity acts as an additional parametric pump that generates additional idlers, and the picture becomes multimode. 

We will first discuss a general situation, and then derive the Bogoliubov transformation coefficients within the balanced mode  model. The results will directly apply to amplification of noise in the threshold region in the presence of a strong tone. The case of developed parametric oscillation is special, and it will be analyzed separately.   

Suppose the input signal has the form, 
\begin{eqnarray}
B_1(t) = (B_1+ b_1(t)) e^{-i\Delta_S t} \,,
\end{eqnarray}
i.e.~in addition to a stronger signal $B_1$ with detuning $\Delta_S$ from $\omega_1+\delta$, 
a weaker classical signal $b_1(t)$ is applied.
Amplification of the strong component is described by \Eqs{AB}--(\ref{V}) (with  $\Delta_S$ replacing $\Delta$). 

The presence of a weak input component will generate an addition to the intracavity field, $(A_n + a_n(t))e^{\mp i\Delta_S t}$, $a_n \ll A_n $, which will be described with the linearized equations,
\begin{eqnarray}\label{eq:EOM_linear}
&& i \dot a_1 + \Delta_S a_1 + \overline\zeta_{1} a_1+ i \Gamma_1 a_1 + \overline\epsilon a_2^\ast \nonumber\\
&& + 2\alpha A_{1}A_{2}^\ast a_2 + \alpha_1 A_{1}^2 a_1^\ast = \sqrt{2\Gamma_{10}} \, b_1(t) \nonumber\\
&& i \dot a_2 -\Delta_S a_2 + \overline\zeta_{2} a_2+ i \Gamma_2 a_2 +  \overline\epsilon a_1^\ast \nonumber\\ 
&& + 2\alpha A_{1}^\ast A_{2} a_1 + \alpha_2 A_{2}^2 a_2^\ast  = \sqrt{2\Gamma_{20}} \, b_2(t) \,.
\end{eqnarray}
Here we introduced the renormalized $\zeta$-coefficients,
\begin{eqnarray}\label{eq:zetas_abovethresh}
\overline\zeta_{1} &=& \delta +  2 \alpha_1 |A_{1}|^2 + 2 \alpha |A_{2}|^2 \nonumber\\
\overline\zeta_{2} &=& \delta +  2 \alpha_2 |A_{2}|^2 + 2 \alpha |A_{1}|^2 \,, 
\end{eqnarray}
and the renormalized pump strength,
\begin{eqnarray}\label{eq:epsilon_abovethresh}
\overline\epsilon &=& \epsilon + 2\alpha A_{1} A_{2}\,.
\end{eqnarray}
A weak input in the second mode, $b_2(t)$, is added for completeness.

In these equations we see a new feature: the amplitude, $a_1$, of the first mode hybridizes 
not only with the conjugated amplitude, $a_2^\ast$, of the second mode as it is in the empty cavity, 
but also with the second mode amplitude, $a_2$,  as well as with its own conjugate, $a_1^\ast$. 
This can be understood as the result of the parametric effect generated by the strong intracavity field with 
frequencies $\omega_{1,2}+\delta$: due to the self-Kerr effect the second harmonics are generated, $2\omega_{1,2}+ 2\delta$, which act as two additional, ``current'' pumps producing a degenerate parametric resonance within each mode \cite{EichlerPRL2011}. 
Furthermore, the cross-Kerr effect generates yet two current pumps  with the combination frequencies, $\omega_2 - \omega_1$ and $\omega_2 + \omega_1$, the former producing parametric frequency conversion, and the latter generates parametric amplification in addition to the generic flux pump. 

\begin{figure}[h]
\includegraphics[width=0.95\columnwidth]{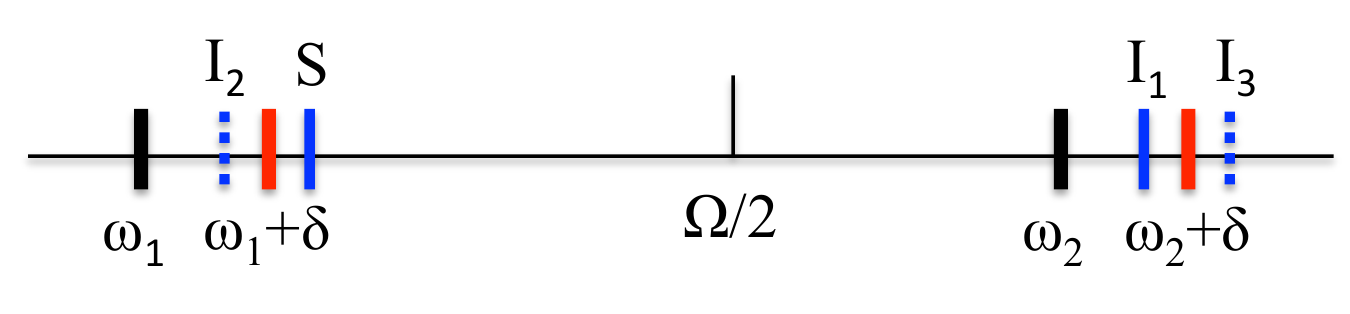}
\caption{
Four-mode structure of the amplified output field for the detuned signal. Black color indicates cavity resonances, 
red color marks parametrically coupled strong field modes (for $\Delta_S=0$); 
solid blue lines indicate signal (S) and primary idler ($I_1$) with frequencies, $\omega_1+\delta+\Delta$ and $\omega_2+\delta -\Delta$; 
dashed blue lines indicate secondary idlers ($I_{2,3}$) with frequencies, $\omega_1+\delta-\Delta$ and $\omega_2+\delta +\Delta$.
}
\label{fig:ModeCoupling}
\end{figure}

The spectrum of the cavity response is illustrated in Fig.~\ref{fig:ModeCoupling}: the weak signal (S) detuned by $\Delta$ from the strong field of the first mode generates a ``primary'' idler (I$_1$) detuned by $-\Delta$ from the strong field of the second mode  as well as two  ``secondary'' idlers (I$_2$, I$_3$) detuned by $\mp\Delta$ from the strong field of respective modes. 

In accord with the structure of \Eq{eq:EOM_linear} we consider a general form of the input field,
\begin{eqnarray}\label{eq:B_abovethresh}
 b_{n}(t) &=&  b_{n}(\Delta) e^{-i \Delta t} + b_{n}(-\Delta) e^{i \Delta t} 
 \;,
\end{eqnarray}
and corresponding form of  the intracavity field, 
\begin{eqnarray}\label{eq:A_abovethresh}
a_n (t) = a_{n}(\Delta) e ^{-i\Delta t} + a_{n}(-\Delta) e ^{i\Delta t} \,.
\end{eqnarray}
The output field, $c_n(t)$, will have the same structure.

It is convenient to write the resulting equations for static amplitudes in matrix form, introducing the two-component vectors 
\begin{eqnarray}
a(\Delta) = 
\left(
\begin{array}{c}
  a_1(\Delta)  \\ a_2(\Delta)
\end{array}
\right)\,, \quad 
b(\Delta) =
\left(
\begin{array}{c}
  b_1(\Delta)  \\ b_2(\Delta)
\end{array}
\right)\,,
\end{eqnarray}
\begin{eqnarray}\label{eq:relation_aDelta_BDelta}
&&{\cal T}(\Delta) a(\Delta) + {\cal E} a^\ast (-\Delta) = {\rm diag}(\sqrt {2\Gamma_{n0}})\, b(\Delta) \\
&&{\cal T}^\ast (-\Delta) a^\ast(-\Delta)  + {\cal E}^\ast a(\Delta) = {\rm diag}(\sqrt {2\Gamma_{n0}})\, b^\ast (-\Delta) \,, \nonumber
\end{eqnarray}
where
\begin{eqnarray}\label{eq:M}
{\cal T}(\Delta) = \left(
\begin{array}{cc}
\Delta + \Delta_S + \bar \zeta_1 + i \Gamma_1  &  2\alpha |A_{1}A_{2} |e^{i\psi}    \\
 2\alpha |A_{1}A_{2} | e^{-i\psi} &   \Delta -\Delta_S + \bar \zeta_2 + i \Gamma_2  
\end{array}
\right) \,,\nonumber\\
\end{eqnarray}
and 
\begin{eqnarray}\label{eq:K}
{\cal E} = \left(
\begin{array}{cc}
 \alpha_1 |A_{1}|^2 e^{2 i\theta_1}&  \overline\epsilon    \\
 \overline\epsilon &     \alpha_2 |A_{2}|^2 e^{2i\theta_2}
\end{array}
\right) \,,
\end{eqnarray}
$\theta_{1,2}$, and $\psi = \theta_1 - \theta_2$, are the phases of the strong intracavity field. 
The matrix  ${\cal T}$ here describes the hybridization of the resonant modes $a_1$ and $a_2$, while the matrix ${\cal E}$ provides the amplification type coupling to the conjugate pair $(a_{1}^\ast, a_{2}^\ast)$.

Inverting \Eq{eq:relation_aDelta_BDelta} we find the intracavity fields and formulate, using the input-output relation, \Eq{eq:cb},   the  Bogoliubov transformation with the matrix coefficients,
\begin{eqnarray}\label{BT2}
c(\Delta) = \hat U(\Delta) b(\Delta) + \hat V(\Delta) b^\ast(-\Delta)
\,.
\end{eqnarray}
These coefficients define the gains for the signal, $G_{11}(\Delta) = |U_{11}(\Delta)|^2$,  and primary idler, $G_{12}(-\Delta) = |V_{21}(-\Delta)|^2$, as well as the gains for secondary idlers,
\begin{eqnarray}
&& G_{11}(-\Delta) = \left|c_1(-\Delta) / b_1(\Delta)\right|^2 = |V_{11}(-\Delta)|^2, \nonumber\\ 
&& G_{12}(\Delta) = \left|c_2(\Delta) / b_1(\Delta)\right|^2 = |U_{21}(\Delta)|^2 \,.
\end{eqnarray}
%

\subsubsection{Balanced mode model}
\label{Sec_analysis_4mode}
To further analyze the multimode amplification process and explicitly evaluate the Bogoliubov coefficients, we apply the methods of multimode squeezing theory \cite{BraunsteinPRA2005,TrepsD2010}. 
This theory operates with a set of ``supermodes'' that diagonalize the matrix Bogoliubov transformation, \Eq{BT2}. 

To do this one needs to diagonalize matrices ${\cal T}$ and ${\cal E}$ in \Eq{eq:relation_aDelta_BDelta}, and   
this can be relatively simply done for the balanced modes. 
In this case the intracavity field amplitudes are approximately equal for strong on-resonance input ($\Delta_S=0$), close to the threshold, \Eq{A1toA2}, $|A_{n}|= |A|$ hence  $\overline\zeta_{n}=\overline\zeta$. 
In the parametric oscillation regime the intracavity field amplitudes are  also equal, \Eq{eq:Asq1_over_Asq2__LC}, and the frequency shift is absent, $\Delta_0=0$.

With these approximations, the matrix ${\cal T}(\Delta)$ is diagonalized by a unitary rotation,
\begin{eqnarray}\label{fraku}
{\mathfrak U} =  {1\over \sqrt 2} \left(
\begin{array}{cc}
 e^{i \psi/2} &  e^{i \psi/2}  \\
e^{-i \psi/2} & -e^{-i \psi/2}  
\end{array}
\right) \,,
\end{eqnarray}
\begin{eqnarray}
&&   {\mathfrak U}^\dag {\cal T}(\Delta) \, {\mathfrak U} 
= {\rm diag}\, 
 (\Delta + \zeta_+ + i\Gamma ,\; \Delta + \zeta_- + i\Gamma) \,, \nonumber\\
&& \zeta_\pm = \overline\zeta \pm 2\alpha |A|^2 = \delta + (4\alpha \pm 2\alpha) |A|^2 \,.
\end{eqnarray}
Matrix ${\cal T}^\ast(-\Delta)$ is diagonalized with the same rotation. Moreover, matrix ${\cal E}$ is also diagonalized with the same rotation,
\begin{eqnarray}
&&{\mathfrak U}^\dag{\cal E}\, {\mathfrak U}^\ast =
 {\rm diag}\,  (\epsilon_+,\,\epsilon_- )\,, \nonumber \\
&& \epsilon_\pm = \pm \overline\epsilon + \alpha |A|^2 e^{i\Theta}
\,. 
\end{eqnarray}
In the supermode basis the field amplitudes take the form, $a_\sigma(\Delta) = {\mathfrak U}_{\sigma n}^\dag a_n(\Delta)$, $\sigma=\pm$,  
and similar for the input and output fields, $b_\sigma$ and $c_\sigma$. 

As the result, \Eq{eq:relation_aDelta_BDelta} splits in the supermode basis in two independent blocks,  
\begin{eqnarray}\label{EOMsigma}
&&\!\! ( \zeta_\sigma + \Delta + i\Gamma)  a_\sigma(\Delta) + \epsilon_\sigma  a_\sigma^\ast(-\Delta)
= \sqrt{2\Gamma} \,  b_\sigma (\Delta)\nonumber\\
&&\!\!(\zeta_\sigma -\Delta  - i\Gamma)  a_\sigma^\ast (-\Delta) + \epsilon_\sigma^\ast a_\sigma(\Delta)
= \sqrt{2\Gamma} \,  b_\sigma^\ast(-\Delta) .
\end{eqnarray}
Here $\Gamma_n = \Gamma_{n0}$ is assumed, neglecting internal losses.

Equations (\ref{EOMsigma}) are similar to the linearized \Eq{AB} for amplification below threshold,
which allows us to readily get the diagonal, two-mode Bogoliubov transformation,
\begin{eqnarray}\label{cbsigma}
 c_\sigma (\Delta) = u_\sigma(\Delta)  b_\sigma (\Delta) + v_\sigma(\Delta)  b_\sigma^\ast(-\Delta) \,, 
\end{eqnarray}
with the Bogoliubov coefficients,
\begin{eqnarray}
\label{eq:uBT_pmbasis}
u_{\sigma}(\Delta) &=& { (\zeta_\sigma +\Delta -  i\Gamma)(  \zeta_\sigma -\Delta   - i\Gamma)    - |\epsilon_\sigma |^2  \over (\zeta_\sigma +\Delta +  i\Gamma)(  \zeta_\sigma -\Delta   - i\Gamma)    - |\epsilon_\sigma |^2 } \nonumber\\
v_{\sigma}(\Delta) &=& {  2i\Gamma\epsilon_\sigma  \over (\zeta_\sigma +\Delta +  i\Gamma)(  \zeta_\sigma -\Delta   - i\Gamma)    - |\epsilon_\sigma |^2} 
\,.
\end{eqnarray}
These coefficients satisfy relations similar to \Eq{uvproperty}, 
\begin{eqnarray}\label{uvsigma_prop}
&& |u_{\sigma}(\Delta)|^2 - | v_{\sigma}(\Delta)|^2  = 1 \nonumber\\
&& u_\sigma(\Delta)v_\sigma(-\Delta)- v_\sigma(\Delta)u_\sigma(-\Delta) = 0 \,.
 \end{eqnarray}

Rotating back to the initial basis, we get the matrix Bogoliubov transformation for the fields of the original modes, \Eq{BT2},
where ${\hat U} = {\mathfrak U} \,{\rm diag} \, (u_{+}, \, u_{-} )\,{\mathfrak U}^\dag$,
\begin{eqnarray}\label{eq:uBT_origbasis}
{\hat U} (\Delta)
= \frac{1}{2}
\left(
\begin{array}{ll}
(u_{+} + u_{-})  &  (u_{+} - u_{-}) e^{i\psi}    \\
 (u_{+} - u_{-}) e^{-i\psi}  &   (u_{+} + u_{-})     
\end{array}
\right)
\,,
\end{eqnarray}
and ${\hat V} = {\mathfrak U}\, {\rm diag} \, (v_{+}, \, v_{-} )\,{\mathfrak U}^T$,
\begin{eqnarray}\label{eq:vBT_origbasis}
{\hat V} (\Delta) 
= \frac{1}{2}
\left(
\begin{array}{ll}
(v_{+} + v_{-}) e^{i\psi}  &  (v_{+} - v_{-})     \\
 (v_{+} - v_{-})    & (v_{+} + v_{-})   e^{-i\psi} 
\end{array}
\right)  \,.
\end{eqnarray}

These equations define amplification gains for all four modes involved. For the   signal and primary idler  we have gains,
\begin{eqnarray}\label{primary}
&& G_{11}(\Delta) = |U_{11}(\Delta)|^2 =  \frac{1}{4} |u_{+}({\Delta}) + u_{-}({\Delta}) |^2 \nonumber\\ 
&& G_{12}(-\Delta) = |V_{21}(-\Delta)|^2 =  \frac{1}{4} |v_{+}({-\Delta}) - v_{-}({-\Delta})|^2 , \;\;
\end{eqnarray}
and for the secondary idlers,
\begin{eqnarray} \label{secondary}
&& G_{11}(-\Delta) = |V_{11}(-\Delta)|^2 = \frac{1}{4} |v_{+}({-\Delta}) + v_{-}({-\Delta})|^2 \nonumber\\ 
&& G_{12}(\Delta) = |U_{21}(\Delta)|^2 =  \frac{1}{4} |u_{+}({\Delta}) - u_{-}({\Delta})|^2 \,.
\end{eqnarray}
Analyzing these equations with help of \Eq{eq:uBT_pmbasis}, one can find that the secondary gains are directly proportional to the intracavity field intensity, $|A|^2$, while the gains (\ref{primary}) persist in the limit, $|A|^2=0$.

\begin{figure}[tbh]
\includegraphics[width=\columnwidth]{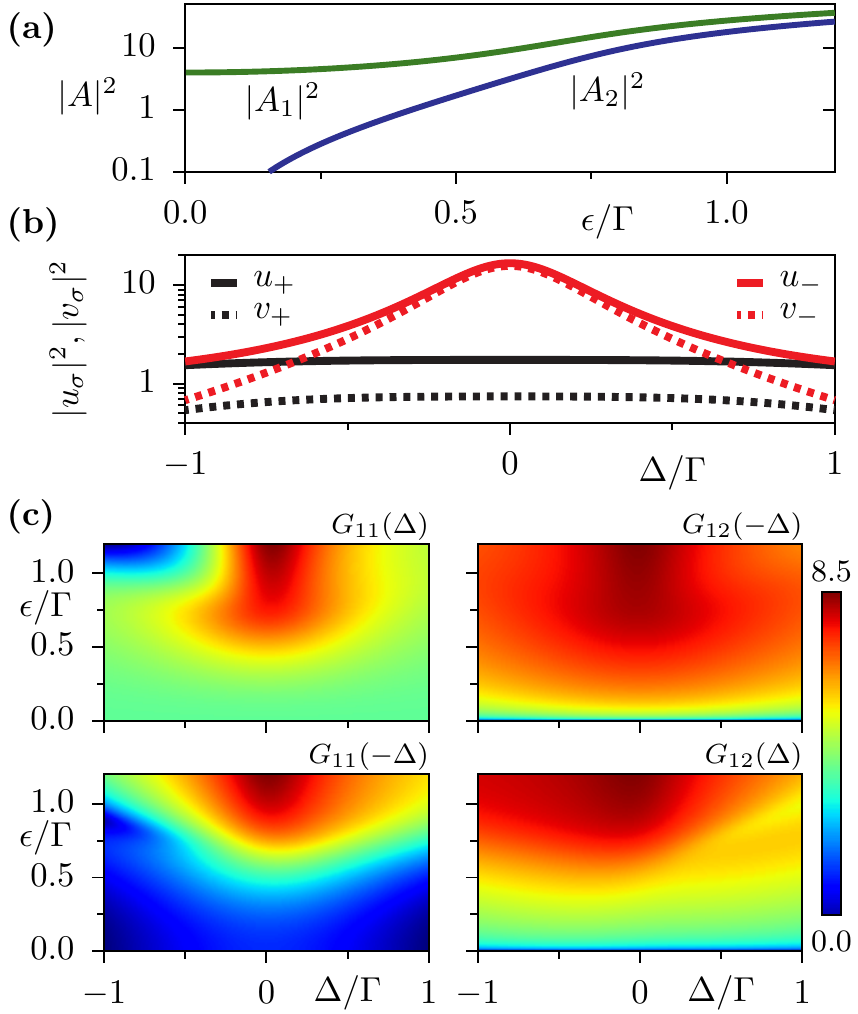}
\caption{
Linear gain spectra for balanced modes
obtained from \Eqs{eq:relation_aDelta_BDelta}--\eqref{BT2} in presence of strong on-resonance input tone 
$|B_1|^2 = 2\Gamma$, $\theta_{B} = 0$.
(a) Cavity field amplitudes $|A_{1,2}|^2$ generated by $B_1$ vs. pump strength $\epsilon$.
(b) Bogoliubov coefficients $|u_\sigma|^2, |v_\sigma|^2$ of supermodes vs. detuning $\Delta$
for $\epsilon = 0.95\Gamma$, computed from \Eq{eq:uBT_pmbasis} using approximation,
$|A|^2 = (|A_1|^2 + |A_2|^2)/2$.
(c) Linear gains $G_{11}(\Delta)$,  $G_{12}(-\Delta)$, $G_{11}(-\Delta)$,  $G_{12}(\Delta)$
vs. input signal detuning $\Delta$ and pump strength $\epsilon$, 
for additional weak signal $b_1(\Delta)$ detuned by $\Delta$ from $\omega_1$. 
($\delta=0$,
$\Gamma_{2}=\Gamma_{1}$, 
$\alpha_2=\alpha_1 = \sqrt{\Gamma_{1}\Gamma_{2}}/100$,
$\Gamma_n=\Gamma_{n0}$.)
}
\label{fig:lineargain_dw1_h_symmodes_withBsq1}
\end{figure}

Applying the obtained formulas to the case of strong input close to the parametric threshold,
\Eqs{A@threshold}, (\ref{A1A2symmetric}), we find the  Bogoliubov coefficients of supermodes (for $\delta=0$), 
\begin{eqnarray}\label{uv_sigma@threshold}
u_\sigma(\Delta)  \approx {- 2\Gamma^2 +\varsigma_\sigma^2  - \Delta^2    \over
\varsigma^2_\sigma - \Delta^2  - 2i\Gamma\Delta } \nonumber\\
v_\sigma(\Delta)  \approx {2 i \sigma \Gamma^2\over 
 \varsigma_\sigma^2 - \Delta^2  - 2i\Gamma\Delta }
 \,,
\end{eqnarray}
with $\varsigma_+ =  \sqrt3 \zeta$, and $\varsigma_-  =  \zeta/\sqrt3$,  $\zeta = 3\alpha|A|^2$. These equations show that amplification of the ``$-$''  
supermode is more efficient, as illustrated in  
Fig.~\ref{fig:lineargain_dw1_h_symmodes_withBsq1}(b).

The linear gain spectra for a detuned input tone $b_1(\Delta)$ 
in presence of a dominant on-resonance input  $B_1$ are illustrated in Fig.~\ref{fig:lineargain_dw1_h_symmodes_withBsq1}(c).
The amplitudes $|A_{1,2}|^2$ of the cavity field generated by $B_1$ are shown in  Fig.~\ref{fig:lineargain_dw1_h_symmodes_withBsq1}(a)
and used in \Eqs{eq:relation_aDelta_BDelta}--\eqref{BT2} to compute the linear gains.
Around the threshold, $\epsilon \approx \Gamma$, the cavity amplitudes are approximately equal, $|A_1|^2 \approx |A_2|^2$, and fulfil the conditions of \Eq{zeta<Gamma}.

\subsection{Four-mode linear amplification in oscillator regime}\label{sec:linamp_abovethresh}

Quite a different situation occurs in the parametric oscillator regime above the threshold. 
Given equations for the oscillation intensity and phase, \Eqs{eq:Asq1_over_Asq2__LC}--\eqref{eq:Theta}, 
we compute the determinant of equation (\ref{EOMsigma}) for the supermodes, 
\begin{eqnarray}\label{Det>threshold}
{\rm Det}_{+} (\Delta) &=& 4(\epsilon^2 - \delta\sqrt{\epsilon^2 - \Gamma^2} - \Gamma^2) -\Delta(\Delta + 2i\Gamma) \,,
\nonumber\\
{\rm Det}_{-} (\Delta) &=&  - \Delta(\Delta + 2i\Gamma) \,.
\end{eqnarray}
Here we see that while ${\rm Det}_+(\Delta)$ is finite everywhere except of the threshold point, 
${\rm Det}_{-}(\Delta=0) $ turns to zero for all pump strengths  above the threshold. 
Consequently, the Bogoliubov coefficients of the ``$-$'' supermode are singular at the oscillation frequency, 
$\Delta=0$. 
Furthermore, they grow with increasing pump strength, 
\begin{eqnarray}\label{u-diverges}
u_-(\Delta) \approx v_-(\Delta) \approx {(4/3)i\Gamma\epsilon \over \Delta(\Delta + 2i\Gamma)}, \quad \epsilon\gg \Gamma \,.
\end{eqnarray}
At the same time the regular coefficients for the ``$+$'' supermode behave qualitatively similar to the degenerate resonance case
\cite{WustmannPRB2013}, 
\begin{eqnarray} 
u_+ \approx 1,  \quad v_+ \approx -{\Gamma^2\over 2\epsilon^2}, \quad \epsilon\gg \Gamma  \,.
\end{eqnarray}

The anomalous behavior of the ``$-$'' supermode implies that it solely defines all the gains 
close to the oscillation frequency, and well above the threshold at all frequencies. In the latter region  
all the four gains have approximately  the same asymptotic scaling,
\begin{eqnarray}\label{G>threshold}
\!\!\! G_{11}(\pm\Delta) \approx   G_{12}(\pm\Delta)   \approx {(4/9)\Gamma^2\epsilon^2 \over \Delta^2(\Delta^2+ 4\Gamma^2)} , \; \epsilon\gg \Gamma \,.
\end{eqnarray}

\begin{figure}[tb]
\includegraphics[width=\columnwidth]{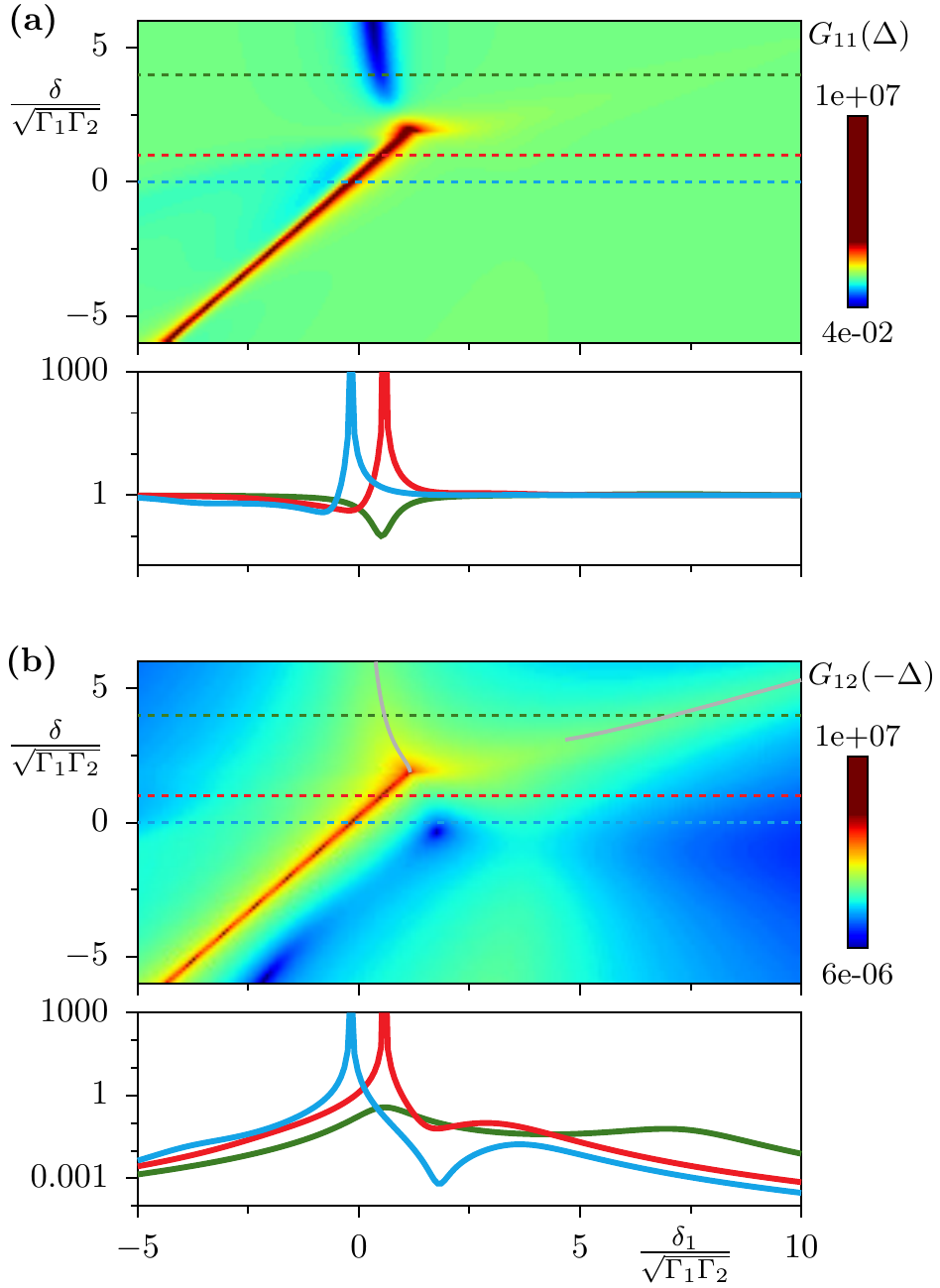}
\caption{
Linear gain spectra for signal and primary idler above threshold, obtained from \Eqs{eq:relation_aDelta_BDelta}--\eqref{BT2}.
(a) Signal gain
$G_{11}(\Delta)$ and  (b) primary idler gain $G_{12}(-\Delta)$ 
vs. input signal detuning $\delta_1$ and pump detuning $\delta$; horizontal dashed lines in color plots indicate cuts presented in lower panels, thin grey lines in (b) indicate positions of resonant peaks. 
Bright region indicates resonant peak at the radiation frequency, $\Delta_0(\delta)$.
Blue strip outside parametric oscillation region, $\delta>\deltath \approx 2\Gmean$,
in (a) is due to internal losses, blue strip within parametric oscillation region in (b) is due to energy transfer to secondary idlers, 
see Fig.~\ref{fig:nontriviallineargain_dw1_dpD_abovethresh}.
($\epsilon=2\sqrt{\Gamma_1\Gamma_2}$,
$\Gamma_{20}=3\Gamma_{10}$, 
$\Gamma_{1} = 1.8\Gamma_{10}$,
$\Gamma_{2} = 4\Gamma_{10} = (4/3) \Gamma_{20}$,
$\alpha_2=3\alpha_1$,
$\alpha_1 = \sqrt{\Gamma_{10}\Gamma_{20}}/100$.)
}
\label{fig:lineargain_dw1_dpD_abovethresh}
\end{figure}

\begin{figure}[tbh]
\includegraphics[width=\columnwidth]{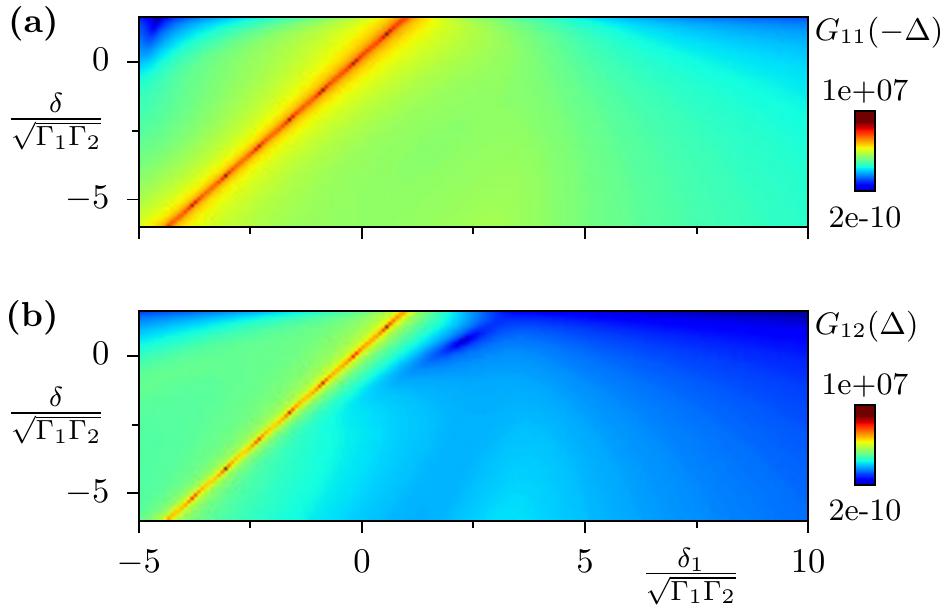}
\caption{
Linear gain spectra for secondary idlers above threshold, (a) $G_{11}(-\Delta)$, and (b) 
$G_{12}(\Delta)$; these idlers appear at  $\delta<\deltath \approx 2\Gmean$, 
in addition to the signal and primary idler presented in Fig.~\ref{fig:lineargain_dw1_dpD_abovethresh}.
The parameters are the same as in Fig.~\ref{fig:lineargain_dw1_dpD_abovethresh}.}
\label{fig:nontriviallineargain_dw1_dpD_abovethresh}
\end{figure}

In Figs.~\ref{fig:lineargain_dw1_dpD_abovethresh} and \ref{fig:nontriviallineargain_dw1_dpD_abovethresh}
the linear gain spectra above the threshold
are shown for different pump and input signal detunings. The plots are made using general equations (\ref{eq:relation_aDelta_BDelta})--(\ref{BT2}) to illustrate the behavior of unbalanced modes and in presence of internal damping.
We remind that $\Delta$ in the parametric oscillation regime, $\delta < \deltath$, refers  to the deviation from the frequency of the oscillation, $\Delta_0$, \Eq{eq:Delta0}. For the signal detuning we again use notation, $\delta_1 = \omega_s - \omega_1$ rather than $\Delta$.

Fig.~\ref{fig:lineargain_dw1_dpD_abovethresh} illustrates the linear gain spectra
of the signal, $G_{11}(\Delta)$, and the primary idler, $G_{12}(-\Delta)$. 
Outside the parametric oscillation region, 
$\delta > \deltath = 1.9 \sqrt{\Gamma_1 \Gamma_2}$, the cavity is empty 
and the detuning $\Delta$ here refers to the reference frequency $\omega_1 + \delta$,
corresponding to the input detuning, $\delta_1 =  \delta + \Delta$. 
The primary idler has the frequency, $\omega_2 + 2\delta-\delta_1$.

The spectrum has similarities with that discussed and shown in Fig.~\ref{fig:lineargain_dw1_dpD_belowthresh},
where now $\deltath$ plays the role of $\delta=0$ in the latter.
For $\delta > \deltath$, the spectrum $G_{12}(-\Delta)$ has two resonance peaks (green curve in lower panel of Fig.~\ref{fig:lineargain_dw1_dpD_belowthresh}(b)). One of these is accompanied by a dip in 
$G_{11}(\Delta)$ stemming from the internal losses. 
When approaching $\deltath$ this dip gradually becomes shallower due to the proximity of the parametric oscillation threshold, while one of the resonance peaks of $G_{12}(-\Delta)$  disappears.  

For balanced modes,  where $\Delta_0=0$,
these resonance positions are similar to the degenerate resonance~\cite{WustmannPRB2013}:
a single resonance at $\delta_1 = \delta$, i.e.~$\Delta=0$, for 
$\deltath \leq \delta \leq \sqrt{\epsilon^2 + \Gamma^2}$,
which splits into $\delta_1 = \delta \pm \sqrt{\delta^2 - \epsilon^2 - \Gamma^2}$ 
at $\delta =\sqrt{\epsilon^2 + \Gamma^2}$.

At $\delta=\deltath$ the parametric resonance lies at 
$\delta_1 = \deltath (1 + (\Gamma_1-\Gamma_2)/(\Gamma_1+\Gamma_2))$,
which coincides there with the parametric oscillation detuning at threshold, 
$\delta_1 = \deltath + \Delta_0(\deltath)$.
Due to the proximity to the parametric oscillation state the resonance is very strong. 

Within the parametric oscillation region, $\delta < \deltath$, the reference frequencies shift to, 
$\omega_1 + \delta + \Delta_0$ and $\omega_2 + \delta - \Delta_0$, respectively. 
Therefore the relation of the input detuning $\delta_1$
and $\Delta$ becomes, $\delta_1 = \delta + \Delta_0 + \Delta$, and the idler frequency is, $\omega_2 + 2\delta-\delta_1$. 
The gains of signal and idler show strong resonance where the input frequency coincides with the frequency of parametric oscillation, $\delta_1 = \delta + \Delta_0(\delta)$, in the whole parametric oscillation region. This agrees with the analytical result derived for the balanced modes, \Eq{G>threshold}. 
Such behavior is drastically different from the degenerate case where the linear gain diverges only at the threshold.

The behavior of the secondary idler gains, $G_{11}(-\Delta)$ and $G_{12}(\Delta)$, are illustrated in  Fig.~\ref{fig:nontriviallineargain_dw1_dpD_abovethresh}. 
These idlers are measured at frequencies,
$\omega_1 + 2\delta - \delta_1 + 2\Delta_0$,
and $\omega_2 + \delta_1 - 2\Delta_0$, respectively.
Due to the distribution of the input power over the four output modes,
the spectral characteristics for each of the idlers become more complicated than for the conventional two-mode amplification below threshold, such as non-monotonous behaviour away from the resonance, compare e.g. $G_{12}(-\Delta)$ at $\delta=0$ (blue curve) in Fig.~\ref{fig:lineargain_dw1_dpD_abovethresh}.

\begin{figure}[tbh]
\includegraphics[width=\columnwidth]{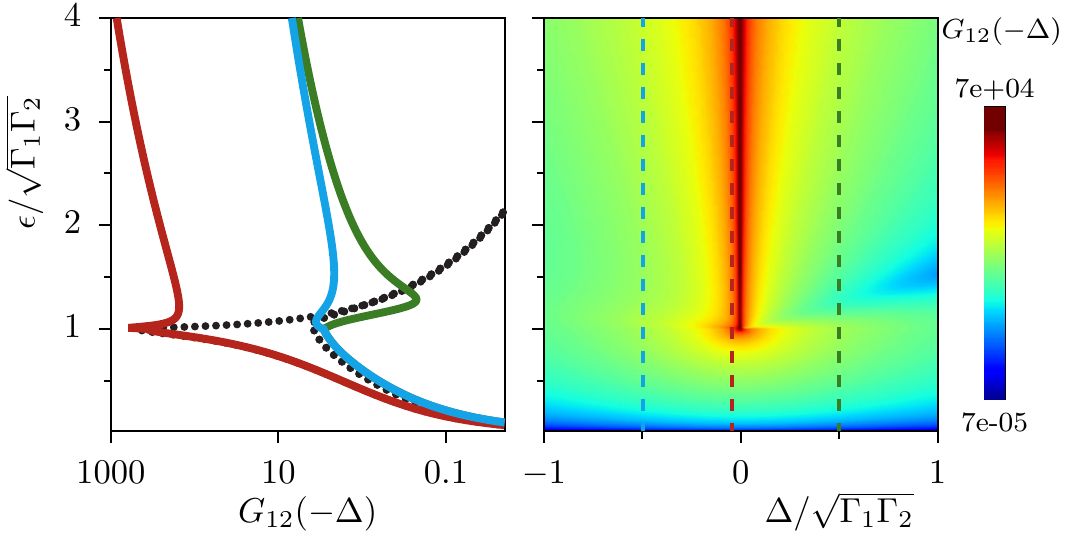}
\caption{
 Linear  idler gain  $G_{12}(-\Delta)$ vs. input signal detuning $\Delta$ and pump strength $\epsilon$,
obtained from \Eqs{eq:relation_aDelta_BDelta}--\eqref{BT2}.
Gain divergence above threshold occurs when input is in resonance with the parametric oscillation, 
$\Delta = 0$. The left panel shows cuts indicated by dashed lines on the right panel.
($\delta=0$,
$\Gamma_{2}=3\Gamma_{1}$, 
$\alpha_2=3\alpha_1 = 3\sqrt{\Gamma_{1}\Gamma_{2}}/100$,
$\Gamma_n=\Gamma_{n0}$.) For a comparison, the idler gain $G_{12}(-\Delta)$
for a degenerate amplifier is presented with dotted lines (parameters  
$\alpha = 3\sqrt{\alpha_1\alpha_2}$, 
$\Gamma = \sqrt{\Gamma_{1} \Gamma_{2}}$).
}
\label{fig:lineargain_dw1_h}
\end{figure}

Fig.~\ref{fig:lineargain_dw1_h} illustrates the linear gain spectrum of the primary idler, $G_{12}(-\Delta)$,
as function of the pump strength $\epsilon$ and input detuning.
It is calculated from \Eqs{eq:relation_aDelta_BDelta}--\eqref{BT2}, where  
the parametric oscillation amplitudes $A_{10}$ and $A_{20}$ are taken into account
above threshold,
$\epsilon >  \sqrt{\Gamma_1 \Gamma_2}$,
while below threshold, with $A_n=0$, the linear gains are identical to those from \Eqs{CB}--\eqref{uv}.
Above threshold
the gain diverges when the input is in resonance  with the parametric oscillation for all pump strengths,
$\delta_1 = \delta + \Delta_0(\epsilon)$, in accord with Figs.~\ref{fig:lineargain_dw1_dpD_abovethresh} and \ref{fig:nontriviallineargain_dw1_dpD_abovethresh}.
Another interesting feature is that the gain does not decrease to zero with 
increasing pump strength, as it is in the degenerate case. In fact all the gains increase with growing 
$\epsilon$, in agreement with \Eq{G>threshold}.
%

\subsubsection{Phase locking and regularization of gain divergence } 
\label{Phase locking}
The divergence of the linear response in the oscillation regime found in \Eqs{Det>threshold}--(\ref{u-diverges}) is a hallmark of non-degenerate parametric resonance \cite{GrahamZPh1968,FabreJdP1989,BjorkPRA1988,DrummondPRA1989}. The divergence is related to  the degeneracy of the free oscillation state with respect to the mode relative phase. 
Indeed, the divergence is caused by the zero value of the determinant of  the supermode dynamical matrix, which implies that this matrix has at least one zero eigenvalue. On the other hand, the zero eigenvalue of the linearized dynamical matrix  indicates an indifferent equilibrium of the oscillator state. 

To eliminate the divergence one needs to go beyond the linear approximation, which can be done for on-resonance input at $\Delta=0$ (we remind that the nonlinear response at finite detuning is non-stationary). 

Our strategy will be to first transform the original nonlinear equations, (\ref{AB}),  to the supermode basis, and then linearize the  equations with respect to $a_+$ while keeping nonlinear terms in $a_-$. 

In terms of the supermode amplitudes for balanced modes, \Eq{AB} takes the form,
\begin{eqnarray}
 \left[{3\alpha\over 2} |A_\sigma|^2   + \alpha |A_{-\sigma}|^2 + i\Gamma \right] A_\sigma 
+ (\sigma\epsilon - {\alpha\over 2}A_{-\sigma}^2) A_\sigma^\ast  \nonumber\\
= \sqrt{2\Gamma} \,b_\sigma \,,  \;\;\;
\end{eqnarray}
where we introduced notation, $b_\sigma = {\mathfrak U}^\dag_{\sigma n} B_n$, and assumed for simplicity $\delta=0$ and $\Gamma=\Gamma_{0}$.

The free solution is nontrivial only for  the ``$+$'' supermode,   
\begin{eqnarray}
A_{+0} &=& \sqrt2 \,|A_0|e^{i\Theta_0/2}, \quad A_{-0} = 0  \,.
\end{eqnarray}
By separating the free solution, $A_{\sigma} = A_{\sigma 0} +  a_\sigma$, and linearizing  the equation for $\sigma =+$ we reproduce the first equation in \Eq{EOMsigma}.
Linearization of the equation for $\sigma=-$ with respect to $a_+$ yields a nonlinear extension of the second equation in \Eq{EOMsigma},
\begin{eqnarray}
&&  \left[{3\alpha\over 2} | a_-|^2  + \zeta_- + i\Gamma \right]  a_-  + \epsilon_-  a_-^\ast = \sqrt{2\Gamma} \, b_- 
\,. 
\end{eqnarray}
It is convenient to introduce for brevity the quantity,
\begin{eqnarray}\label{eq:a-}
 Q = \zeta_- + i\Gamma = {2\over 3}\sqrt{\epsilon^2-\Gamma^2} + i\Gamma \,,
 \end{eqnarray}
and variables $\bar a_- = a_- e^{-i\Theta_0/2}$ and $\bar b_- = b_- e^{-i\Theta_0/2}$, then the equation takes the form,
\begin{eqnarray}\label{eq:a-}
 {3\alpha\over 2} |\bar a_-|^2 \bar a_- +  Q  \bar a_-  + Q \bar a_-^\ast = \sqrt{2\Gamma} \,\bar b_- \,.
 \end{eqnarray}
Following the standard way of inverting this equation and substituting the solution into the input-output relation, we obtain the nonlinear Bogoliubov transformation,
\begin{eqnarray}\label{c-}
\bar c_- &=& {\left({3\alpha\over 2} |\bar a_-|^2  + Q^\ast \right)^2 - |Q |^2 \over {\rm Det}} \,\bar b_- 
+ {2i\Gamma Q   \over {\rm Det}} \,\bar b_-^\ast \\
&=& \bar u_- \bar b_- + \bar v_- \bar b_-^\ast \nonumber
\end{eqnarray}
\begin{eqnarray}
{\rm Det} &=& 
 {3\alpha\over 2} |a_-|^2 \, \left({3\alpha\over 2} |a_-|^2  + Q + Q^\ast \right) \,.
\end{eqnarray}
One can straightforwardly check that the relation between the nonlinear Bogoliubov coefficients, $|\bar u_-|^2 - |\bar v_-|^2 = 1$, still holds.

Here we see that even in the limit $\bar a_- \to 0$ the determinant remains finite for $\epsilon > \Gamma$, in contrast to the determinant of \Eq{EOMsigma} for the ``$-$'' supermode at $\Delta=0$.
Only at the threshold, where $Q=i\Gamma$, the determinant can turn to zero.  Summarizing, a  more accurate calculation which includes the nonlinearity results in a regular response to the on-resonance input. 

This result however imposes a problem. The ``input'' amplitude, 
$\bar b_- = (e^{-i\psi/2} b_1 - e^{i\psi/2} b_2) e^{-i\Theta_0/2}/\sqrt{2}$,  
in \Eq{c-} contains the uncertain phase $\psi$, which is also present in the Bogoliubov coefficients for the original modes, \Eq{eq:uBT_origbasis}. This phase was treated so far as a free parameter, but in  fact it is locked in the driven oscillator as it was mentioned in Section~\ref{onresonance_balanced}.

To reveal the locking effect we make a polar decomposition of the total intracavity fields, $A_{1,2} = r_{1,2} e^{i(\Theta \pm \psi)/2}$, then the supermode amplitudes read,
\begin{eqnarray}
A_\pm = {r_1 \pm r_2 \over \sqrt 2} \, e^{i\Theta/2} \,.
\end{eqnarray}
Using this representation we are able to formulate the constraint, ${\rm Im} \,(A_-/ A_+) = 0$,
which reads for $\bar a_-$,
\begin{eqnarray}
 {\rm Im}\, {\bar a_-\over \sqrt2 |A_0|+  \bar a_+} = 0\,,
 \end{eqnarray}
with $\bar a_+ = a_+ e^{-i\Theta_0/2}$, or
\begin{eqnarray}\label{constraint}
\sqrt2 |A_0| \,{\rm Im}\, \bar a_-   + {\rm Im}\,( \bar a_- \bar a_+^\ast) = 0 \,.
 \end{eqnarray}
This nonlinear equation establishes a missing relation between the oscillation relative phase and the  on-resonance input amplitudes.   

Having in hand \Eq{constraint} we are now able to construct a consistent solution to  \Eq{eq:a-} in the linear approximation. In this limit \Eq{constraint} reduces to ${\rm Im}\,\bar a_- =0$, which yields the linearized solution, 
\begin{eqnarray} \label{a-}
\bar a_- = {\rm Re} \,\bar a_-  = \sqrt{\Gamma\over 2}\, {\bar b_- \over Q}  \,,
\end{eqnarray}
and the constraint on the source, ${\rm Im} \,(Q^\ast \bar b_- ) = 0$, or
\begin{eqnarray}\label{constraint_-}
Q^\ast  b_- = Q e^{i\Theta_0} b_-^\ast \,.
\end{eqnarray}
One can explicitely extract the phase factor from this equation, and express it through the mode  inputs, e.g.~for the single mode input, $B_2=0$,
\begin{eqnarray} \label{eq_psi}
e^{i\psi} =  e^{-i\Theta_0} \,{Q^\ast\over Q} \, {B_1\over B_1^\ast} \,.
\end{eqnarray}
Thus the uncertain relative phase of the free oscillator is locked by the on-resonant input and defined by the phase of this input.

Due to the phase-locking effect, the on-resonance output field becomes regular,
\begin{eqnarray}\label{}
c_- &=& { Q+Q^\ast  \over 2Q  } \,b_- \,,
\end{eqnarray}
in the linear approximation.

In fact the phase locking can also be used to eliminate the divergence of the linear response to a detuned input  at 
$\Delta \to 0$. 
To this end we note that the singular gains in \Eq{G>threshold} were defined for a single harmonic input, $b_-(\Delta)$, at detuning $\Delta$, while $b_-^\ast(-\Delta)$ at detuning $-\Delta$ is assumed zero (cf.~gain definitions at finite detuning, e.g.~\Eqs{primary}, \eqref{secondary}). However, at zero detuning a regular Bogoliubov transformation results from the interference of the input amplitudes, $b_-(0)$, and $b_-^\ast(0)$.  

Taking this remark into account, we consider a slightly broadened input, whose spectrum contains contributions of  both positive and negative $\Delta$-harmonics. 
Expressing the Bogoliubov transformation in \Eqs{cbsigma}--(\ref{eq:uBT_pmbasis}) through $Q$, 
\begin{eqnarray}\label{}
c_-(\Delta) =  {2i\Gamma \; 
[ Q^\ast b_- (\Delta)  - Qe^{i\Theta_0} b_-^\ast (-\Delta) ] + \Delta^2b_- (\Delta) \over \Delta(\Delta+2i\Gamma )} \,, \nonumber
\end{eqnarray}
we insert \Eq{constraint_-} into the square brackets to get,
\begin{eqnarray} 
Q^\ast (b_- (\Delta) - b_-(0))  - Qe^{i\Theta_0} (b_-^\ast (-\Delta)-b_-^\ast(0)) \,. \nonumber
 \end{eqnarray}
Assuming  a spectrally smooth input, $b_- (\Delta) - b_-(0) \approx  b'_-(0) \Delta$, we arrive at the output field, 
\begin{eqnarray} 
c_- (\Delta) \approx {2i\Gamma [ Q^\ast b'_-(0) + Q e^{i\Theta} b_-'^\ast (0)] + 
\Delta \, b_- (\Delta) \over 2i\Gamma + \Delta} \,, \nonumber
\end{eqnarray}
that is regular at $\Delta =0$.


\section{Quantum fluctuations}
\label{Fluctuation}

In this section we consider quantum fluctuations of the output field of the non-degenerate parametric amplifier. 
We will use the results of the classical analysis in Section~\ref{Amplification}  to compute the noise spectral densities and evaluate the effect of the noise squeezing and output signal to noise ratio.

The amplifier output  consists of the sum of the classical signals studied so far, $C_n(t)$, and the quantum noise represented by the bosonic operators, $c_n(t)$, in both parametrically coupled modes. The output is concentrated around the  frequencies of respective modes, $\omega_n+\delta$, within  bandwidths, $\Gamma_{n}$. 
The measurement of the output microwave signals are performed via homodyne detection \cite{YurkeBook}, when the output fields are mixed with the fields of two local oscillators, $A_{LO}e^{i(\omega_n+\delta)t + i\theta_n} + {\rm c.c.}$, and low frequency envelopes are filtered out yielding  quadratures,
\begin{eqnarray}\label{eq:quadrature}
X(t) &=& \sum_{n=1,2} [X_{n}^{\theta_n}(t) + x_n^{\theta_n} (t)]  \nonumber\\
& = & \sum_{n} \, [(C_{n}(t) +  c_{n}(t) ) \,e^{-i\theta_n} + {\rm h.c.}] \,
\end{eqnarray}
(in this section we use the notation, $\theta_{1,2}$, for the local oscillator phases rather than the phases of the classical oscillator amplitudes, which will not be discussed here). 
The spectral power for the obtained quadratures is defined as, 
\begin{eqnarray}\label{P}
P(\Delta) = \lim_{T\to\infty} \,{1\over 2T}\left| \int _{-T}^{T} dt \,X(t)  e^{i\Delta t} \right|^2  = P_0(\Delta) + S(\Delta) \,,\nonumber\\
\end{eqnarray}
where $P_0(\Delta)$ refers to the coherent signal component, and reads for a single  input tone with frequency, $\omega_1+\delta$, 
\begin{eqnarray}\label{P0}
P_0(\Delta) = 2\pi\delta(\Delta) \left[\sum_n \left(C_n e^{-i\theta_n} + C_n^\ast e^{i\theta_n}\right) \right]^2 . 
\end{eqnarray}
The second term in \Eq{P} refers to the noise component and is commonly quantified with the squeezing spectral density\cite{SqueezingBook},
\begin{eqnarray}\label{S}
\!\!\!  S(\Delta) = \int_{-\infty}^{\infty} dt \, e^{i\Delta t}\la x(t)x(0)\ra = \sum_{nm} S_{nm}^{\theta_n\theta_m}(\Delta).
\end{eqnarray}

Noise squeezing is manifested in the anisotropy of the quadrature spectral density with respect to the local oscillator phases, which results from  the interference between the signal and idler. Noise squeezing in a single mode is not possible in the two-mode linear amplification regime. On the other hand, under the four-mode nonlinear amplification noise squeezing becomes detectable even in the single mode measurement.

\subsection{Squeezing spectral density}
The partial components of the squeezing spectral density in \Eq{S} can be expressed through Fourier harmonics of the noise quadratures,
\begin{eqnarray}\label{Snm}
S_{nm}^{\theta_n\theta_m}(\Delta) = \int _{-\infty}^{\infty}  d\Delta'
\,\langle x_n^{\theta_n}(\Delta) x_m^{\theta_m}(\Delta')\rangle  
= S_{mn}^{\theta_m\theta_n}(\Delta).\nonumber\\
\end{eqnarray}
where
\begin{eqnarray}
x_n^\theta(\Delta) = \!\! \int _{-\infty}^{\infty} \!\! {dt\over \sqrt{2\pi}}\,  e^{i\Delta t}  x^\theta_n (t)  
= c_{n}(\Delta)   e^{-i\theta} + c_{n}^\dag(-\Delta)  e^{i\theta} .\nonumber\\
\end{eqnarray}
In what follows the averaging is assumed over the input vacuum state.   

To compute the correlation functions in \Eq{Snm} we will use the linearized Bogoliubov transformation either in the form of \Eqs{CB}--(\ref{uv}), for the two-mode squeezing,   or in the form, \Eqs{BT2}, (\ref{eq:uBT_origbasis})--(\ref{eq:vBT_origbasis}), for the four-mode squeezing of balanced modes.
In the general case of unbalanced modes, \Eqs{eq:relation_aDelta_BDelta}--\eqref{BT2} have to be used.

Due to the linear form of the Bogoliubov transformation it is readily extended to the quantum case by replacing classical amplitudes with the bosonic operators. It is important that the Bogoliubov transformation preserves the canonical commutation relations, i.e. the output operators are bosonic provided the input operators are bosonic \cite{CavesPRD1982},
\begin{eqnarray}
[c_n(\Delta)\,, c_m^\dag(\Delta')] = [b_n(\Delta)\,, b_m^\dag(\Delta')] =
\delta_{nm}\delta(\Delta - \Delta')\,. \nonumber
 \end{eqnarray}
This property is fulfilled in the absence of internal losses (otherwise one should add additional input noise channels), 
and in the two-mode case it follows from the properties of the $u-v$ coefficients in \Eq{uvproperty}.
In the four-mode case, the supermode operators respect the commutation relation due to similar relations between the $u-v$ coefficents,  \Eq{uvsigma_prop}, and since the supermodes are connected to the original modes  via unitary rotation, the commutation relations hold also for the mode operators (cf. Ref. \cite{BraunsteinPRA2005}). 

The latter argument does not, however, apply to the parametric oscillator regime. In this case, the Bogoliubov coefficients in \Eqs{eq:uBT_origbasis}--(\ref{eq:vBT_origbasis}) depend on the free oscillation phase $\psi$. As we found from the classical analysis, this phase is defined by the input, and therefore it is susceptible to the input noise. Therefore, this phase undergoes quantum fluctuation, which is correlated with the input quantum noise.  As a result, the structure of the output noise  correlation functions must be qualitatively different from the ones in the sub-threshold region, and from conventional two-mode squeezing for the degenerate oscillator \cite{WustmannPRB2013}. This topic requires separate treatment, which goes beyond the scope of this paper. 

\subsection{Weak signal, two-mode noise squeezing }
First we analyse \Eqs{S}--(\ref{Snm}) in the case of two-mode squeezing  for 
amplification of a weak signal below threshold. In this case, the input-output relation, \Eq{CB}, 
with linearized Bogoliubov coefficients, \Eq{uv}, apply independently to both signal and noise.
For equal local oscillator phases, $\theta_1 = \theta_2 = \theta$, we get,
\begin{eqnarray}
 S_{n}^\theta(\Delta)  &=&  |u_n(\Delta)|^2 + |v_n(-\Delta)|^2  \,, \nonumber\\
 S_{12}^\theta(\Delta) &=& u_1(\Delta)v_2(-\Delta) e^{-2i\theta} + 
u^\ast_2(\Delta)v_1^\ast(-\Delta)e^{2i\theta}\,.  \nonumber \\
\label{S2mode_diag_nondiag}
&=&  [S_{21}^\theta(\Delta)]^\ast \,.
\end{eqnarray}

The full spectral density can be transformed, after regrouping using the symmetry properties of the $u-v$ coefficients,  
to the form, 
\begin{eqnarray}\label{S2mode3}
&&S^\theta(\Delta)  \\
&& = \sum_{\pm} \left[ e^{- 2r(\pm\Delta)}
+\;  2\sinh2r(\pm\Delta) \cos^2\left({\chi(\pm\Delta)\over 2}- \theta\right) \right] \,,  \nonumber
\end{eqnarray}
where we introduced the squeezing parameter, $r( \Delta)$, through the relation,
\begin{eqnarray}\label{defr}
|v_1(\Delta)| = |v_2(-\Delta)| = \sinh r( \Delta), 
\end{eqnarray}
and the phase,
\begin{eqnarray}\label{}
\chi(\Delta) = {\rm arg}\,[u_1(\Delta)v_2(-\Delta)] \,.
\end{eqnarray}

\begin{figure}[tb]
\centering
 \includegraphics[width=\columnwidth]{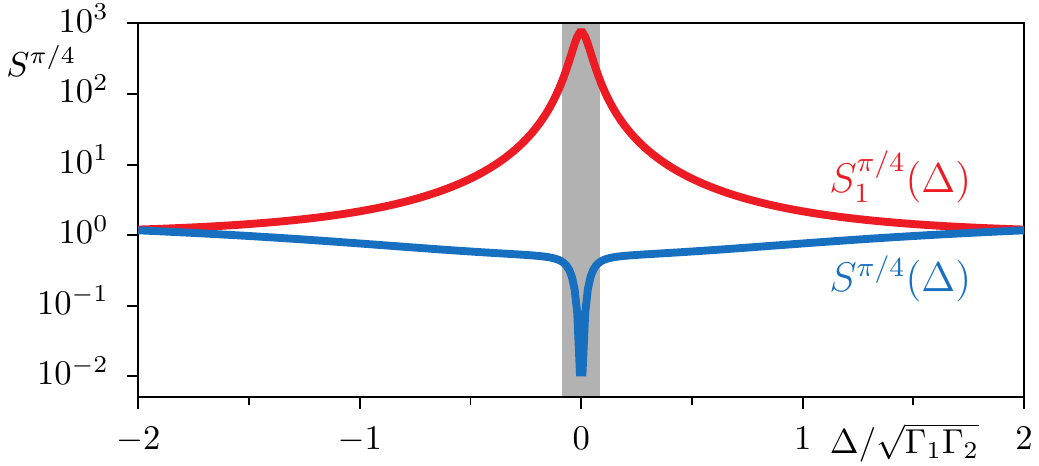}
\caption{
Linear squeezing spectra $S_{1}^{\pi/4}(\Delta)$ and $S^{\pi/4}(\Delta)$ vs. input detuning $\Delta$
for two-mode squeezing in empty cavity.
Individual mode spectral density $S_{1}^{\pi/4}(\Delta)$ from \Eq{S2mode_diag_nondiag} (red).
Full two-mode spectral density 
$S^{\pi/4}(\Delta) = (S_1^{\pi/4} + S_2^{\pi/4} + S_{12}^{\pi/4} + S_{21}^{\pi/4})(\Delta)$ 
from \Eq{S2mode_diag_nondiag} (blue).
The grey shaded area marks the region $|\Delta| < \sqrt{\Gamma_1\Gamma_2} - \epsilon$. The local oscillator phase is chosen in maximum squeezing direction, $\theta=\pi/4$, \Eq{theta0}, using $\chi(0)=-\pi/2$.
($\epsilon=0.95\sqrt{\Gamma_1\Gamma_2}$, $\delta=0$,  
$\Gamma_{2}=3\Gamma_{1}$, 
$\alpha_2=3\alpha_1 = 3\sqrt{\Gamma_{1}\Gamma_{2}}/100$, $\Gamma_n=\Gamma_{n0}$.)
}
\label{Fig:Slinear}
\end{figure}

As one can see from \Eq{S2mode_diag_nondiag}   the output noise  mixed with only one local oscillator, $S_1(\Delta)$, 
does not contain $\theta$-dependent interference terms, implying that the noise of a single mode is not squeezed but just amplified. This results from the large difference in the signal and idler frequencies. In the limit of strong (but still linear) amplification, $G_{11} \approx G_{12} \gg 1$, close to the threshold, $\sqrt{ \Gamma_1\Gamma_2} - \epsilon \ll \Gamma_n$  ($\delta=0$), the spectral density is proportional to the gain,
\begin{eqnarray}\label{S1}
S_{1}(\Delta) \approx 2|u_1(\Delta)|^2 = 2G_{11}(\Delta)\,,
\end{eqnarray}
and has a  sharp peak localized at small $|\Delta|  \lesssim \sqrt{ \Gamma_1\Gamma_2} - \epsilon \ll \Gamma_n$,  shown in Fig.~\ref{Fig:Slinear}. 
 
In order to achieve noise squeezing,  one  needs to employ two local oscillators, in which case the full noise, \Eq{S2mode3}, includes the interference term, and the amplification effect described by this term can  be cancelled by appropriate choice of the local oscillator phase $\theta$, 
\begin{eqnarray}\label{theta0}
\theta ={\pi\over 2} + {\chi(0)\over 2} \,.
\end{eqnarray}
In this maximum squeezing direction the spectral density has a sharp dip, as illustrated in Fig.~\ref{Fig:Slinear}, with a minimum value, $S_{\text{min}}(0) \approx 1/(2G_{11}(0))$, while in the $\pi/2$-shifted maximum amplification direction it has a sharp peak with maximum value,  $S_{\text{max}}(0) \approx 8G_{11}(0)$. 

%

\subsubsection{Signal to noise ratio}
The efficiency of amplification of the signal with respect to the noise is quantified 
with the signal to noise ratio, 
\begin{eqnarray}
\text{SNR} = {{\rm max}\, \overline {P_0^\theta} \over \overline{S^\theta}}\,,
\end{eqnarray}
where bar indicates integration over some bandwidth, $(-\bar\Delta/2, \bar\Delta/2)$. 
 
In the case of the homodyne detection with one local oscillator the signal output amplitude is, $C_1 = u_1(0) B_1 = - \sqrt{G_{11}(0)} B_1$. This gives according to \Eq{P0}, 
\begin{eqnarray}\label{P0bar}
\overline {P_0^\theta} &=& 8\pi G_{11}(0)\cos^2(\theta_{B} - \theta) |B_1|^2 \,.
\end{eqnarray}
Integrating the noise  over a small bandwidth, $\bar\Delta/2 \ll \sqrt{\Gamma_1\Gamma_2} - \epsilon$, we get, $\overline{S_{1}} \approx 2G_{11}(0)\bar\Delta$, and 
\begin{eqnarray}\label{SNRlin}
\text{SNR} &=& 4\pi  \,{|B_1|^2\over \bar\Delta} \,. 
\end{eqnarray}
This is half of the  input SNR value. The factor 1/2 reflects the noise added by the idler. 
 
Homodyne detection with  two local oscillators adds the idler contribution also to the signal output, 
$C_2 = v_2(0) B_1^\ast$, which results in the spectral power,
\begin{eqnarray}\label{P0weak}
\overline{P_0^\theta} &=&  32\pi G_{11}(0) 
\cos^2 \left(\theta_B + {\pi\over 4} \right) \nonumber\\
&\times&\cos^2 \left({\chi(0)\over 2}  - \theta \right)\,|B_1|^2 \, .
\end{eqnarray}
Comparing this result with \Eq{S2mode3} we find that the maximum  squeezing direction for the noise and the signal coincide.  For the maximum amplification direction the signal to noise ratio,
\begin{eqnarray}\label{SNRlin}
\text{SNR} &=& 8\pi  \,{|B_1|^2\over \bar\Delta} \,, 
\end{eqnarray}
is equal to the input SNR. 

The absence of improvement of the signal to noise ratio in the linear amplification regime is explained by the fact that the coefficients of the Bogoliubov transformation  are the same for the noise and for the signal.

\subsection{Strong signal, four-mode noise squeezing}

The situation is different for the nonlinear amplification discussed in Section~\ref{sec:nlin@threshold}, when the  nonlinear frequency shifts define the height and the width of the resonance,  $\alpha_n |A_n|^2 \gg \sqrt{|\Gamma_1\Gamma_2-\epsilon^2|}$. We remind that this regime is valid  across the threshold region including the oscillator region as long as the intracavity field of the signal dominates over the oscillation field. 

In this case, the components of the noise spectral density are expressed through the matrix Bogoliubov coefficients in \Eqs{eq:uBT_origbasis}--\eqref{eq:vBT_origbasis}, and have the form, 
\begin{eqnarray}\label{SUV}
&& S_{nm}^{\theta}(\Delta) = 
U_{nk}(\Delta)U_{mk}^\ast(\Delta) +
 V_{nk}^\ast(-\Delta) V_{mk}(-\Delta)  \nonumber\\
&&+  U_{nk}(\Delta)V_{mk}(-\Delta) e^{-2i\theta} +
V_{nk}^\ast(-\Delta) U_{mk}^\ast(\Delta) e^{2i\theta}\,.  \nonumber\\
&&
\end{eqnarray}
Because of the presence of the secondary idlers one would expect that already the noise of one mode can be squeezed. Using the balanced mode model, we compute  the diagonal component, $S_{11}$, in \Eq{SUV}, 
\begin{eqnarray}\label{S11strong}
&&S_{11}^{\theta}(\Delta) = 
{1\over 2}\sum_{\sigma=\pm} \left\{ (|u_\sigma(\Delta)| - |v_\sigma(-\Delta)| )^2  \right. \\
&&\hspace*{0.5cm} + \left.  2|u_\sigma(\Delta)  v_\sigma(-\Delta)| \,
[1+\cos(\psi+\chi_\sigma(\Delta)- 2\theta)] \right\} \,. \nonumber
\end{eqnarray}
This component consists of independent contributions of both supermodes, where each contains an interference term  similar to \Eq{S2mode3}. 

The optimum squeezing direction for each supermode is defined by the phase,   
$\chi_\sigma(\Delta) = {\rm arg}\,[ u_\sigma(\Delta)  v_\sigma(-\Delta)]$, and these phases are different for different supermodes, 
$\chi_\sigma(0)= -\sigma \pi/2$, as it follows from \Eq{uv_sigma@threshold}. Together with the value, $\psi = 2\theta_B+\pi/2$, that follows from \Eq{A1A2symmetric}, this  gives the interference terms  in \Eq{S11strong}, proportional to $[1 + \sigma\cos 2(\theta_B-\theta)]$.  Therefore both noise supermodes cannot be squeezed simultaneously.

The spectral power of the coherent signal has the form similar to \Eq{P0bar}, with the nonlinear gain (cf. \Eqs{Gnonlinear},(\ref{A@threshold})),
\begin{eqnarray}\label{}
G_{11}(0) \approx  \left({2\Gamma^2\over \zeta^2}\right)^2\,,
\end{eqnarray}
and maximum amplification direction, $\theta=\theta_B$. Comparing this with the noise squeezing directions we find that it coincides with the squeezing direction of the  ``$-$'' supermode, while the 
contribution of the ``$+$'' supermode is amplified. The noise spectral density of the latter component has a sharp peak confined to the interval,  $|\Delta| \lesssim \varsigma_+^2/\Gamma$, according to \Eq{uv_sigma@threshold}, and has the peak value, 
\begin{eqnarray}\label{}
S_{11}^{\theta_B}(0) \approx  2\left({2\Gamma^2\over 3\zeta^2}\right)^2 = {2\over 9}\,G_{11}(0)  \,.
\end{eqnarray}
This leads to the output SNR,
\begin{eqnarray}\label{SNR_nlin_singleLO}
\text{SNR} &=& 36 \pi \,{|B_1|^2\over \bar\Delta} \,, 
\end{eqnarray}
which is nine time larger than the linear result, \Eq{SNRlin}. 
This enhancement of the signal to noise ratio results from the difference between the differential gain, which characterizes the noise, and the nonlinear gain of the signal;  the former is always smaller for the nonlinear amplification.

A detailed numerical study confirms the validity of the adopted approximations for the absolute values of the intracavity fields, and the Bogoliubov coefficients that lead to \Eq{SNR_nlin_singleLO}. However, the numerically evaluated phases of these quantities deviate from the analytical values indicating high sensitivity of the phases to the approximation made. This deviation leads to drastic further enhancement of the SNR.

\begin{figure}[tb]
\centering
 \includegraphics[width=\columnwidth]{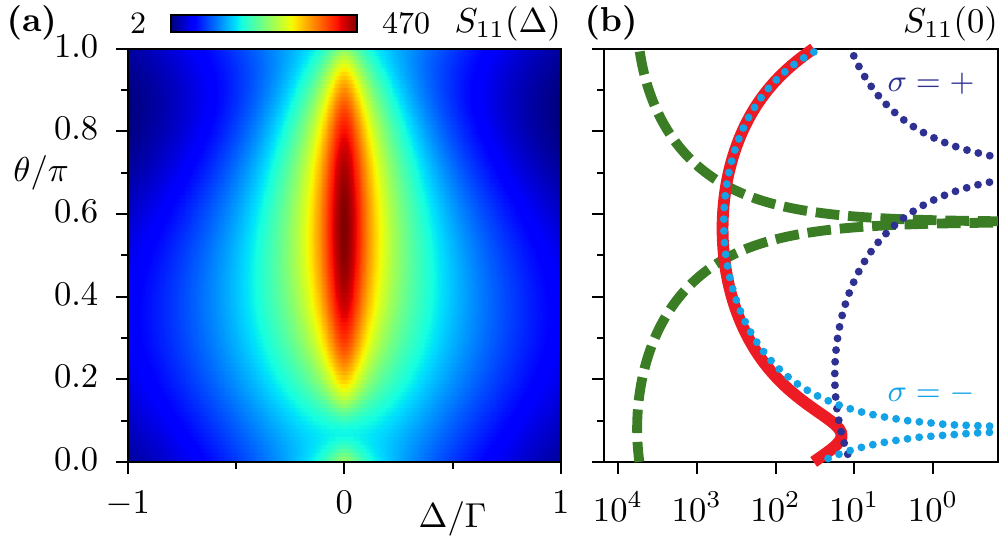}
\caption{
Output noise squeezing in presence of strong on-resonance input $B_1$, mixed with single local oscillator at frequency $\omega_1$ and phase $\theta$:
(a) Noise squeezing spectrum $S_{11}^\theta(\Delta)$, \Eq{SUV}, using linearized matrix Bogoliubov coefficients 
from \Eq{eq:relation_aDelta_BDelta}--\eqref{BT2}.
(b) Noise $S_{11}^\theta(0)$ vs. $\theta$  (red solid line) and its decomposition in supermode contributions, \Eq{S11strong} (blue dashed line and light blue dotted line);  classical quadrature response $\overline{P_0}$, \Eq{P0} , with amplitude $C_{1}$ calculated from \Eqs{eq:Aampl}--\eqref{C} is shown with green dashed line.
($\epsilon=\Gamma, \delta=0,  |B_1|^2 = 0.1\Gamma, \theta_B = 0, 
\Gamma_1=\Gamma_2,
\alpha_1 = \alpha_2 = \Gamma/100$, $\Gamma_n=\Gamma_{n0}$.)
}
\label{fig:S11}
\end{figure}

The result of the numerical study is illustrated in Fig.~\ref{fig:S11} for the representative case of signal input, $|B_1|^2 = 0.1\Gamma$, at the threshold, $\epsilon=\Gamma$. In panel (a) we show the noise spectral density $S_{11}^\theta(\Delta)$ versus $\Delta$ and $\theta$, and in panel (b) compare $S_{11}^\theta(0)$ (red line) with the spectral power of the signal $\overline{P_0^\theta}$ (green dashed line). 
As expected, the squeezing direction of the ``$-$'' supermode (light blue dotted line) approximately coincides with the maximum signal amplification. 
However, the spectrum of the ``$+$'' supermode (dark blue dotted line) is shifted by less than $\pi/2$ from the ``$-$'' supermode spectrum. This results in stronger suppression of the overall noise in the direction of the maximum signal amplification. The maximum SNR value is achieved at $\theta=\theta_B+ 0.06\pi$, where $\overline P_0 \approx 
1820\pi |B_1|^2$ and $S_{11}(0) \approx 14.5$, giving,
\begin{eqnarray}\label{}
\text{SNR} &\approx& 125 \pi \,{|B_1|^2\over \bar\Delta} \,. 
\end{eqnarray}
This is about 30 times larger than the linear result, and 15 times larger than the input value.

We conclude this section with a discussion of the noise squeezing in a homodyne detection with two local oscillators. To this end we compute the cross mode correlation function,
\begin{eqnarray}\label{S12strong}
&& S_{12}^{\theta}(\Delta) = {1\over 2}\sum_{\sigma=\pm} \sigma \,\{ 
   |u_\sigma(\Delta)|^2  e^{i\psi}   
 +  |v_\sigma(-\Delta)|^2   e^{-i\psi}  \nonumber\\
 &&+    \left[ u_\sigma(\Delta)  v_\sigma(-\Delta)    e^{-2i\theta} + {\rm c.c.} \right] \} 
 = [S_{21}^{\theta}(\Delta)]^\ast \,,
 \end{eqnarray}
and collect all the correlation function components to obtain after some algebra,
\begin{eqnarray}\label{Stotalstrong}
&&S^{\theta}(\Delta) = \nonumber\\
&& 2\{ e^{-2r_+}  + \sinh 2r_+ [1 +   \cos (\chi_+ - 2\theta)]  \} \cos^2{\psi\over 2} + \nonumber\\
&&  2\{ e^{-2r_-} + \sinh 2r_- [1- \cos (\chi_- - 2\theta)]  \} \sin^2{\psi\over 2},
\end{eqnarray}
here $|v_{\sigma}(\Delta)| = \sinh r_\sigma(\Delta)$. 

Given the relation, $\chi_\sigma(0) = -\sigma\pi/2$, we find that the optimum squeezing is achieved at $\theta = \pi/4$ regardless of $\psi$ value, when both terms in the square brackets turn to zero. Therefore the full squeezing of noise is possible. At the same time, the spectral power of the strong coherent signal in this case is given by \Eq{P0weak}, with $\chi(0) = -\pi/2$, and  the signal turns to zero for the noise squeezing direction, $\theta= \pi/4$. 

\begin{figure}[tb]
\centering
 \includegraphics[width=\columnwidth]{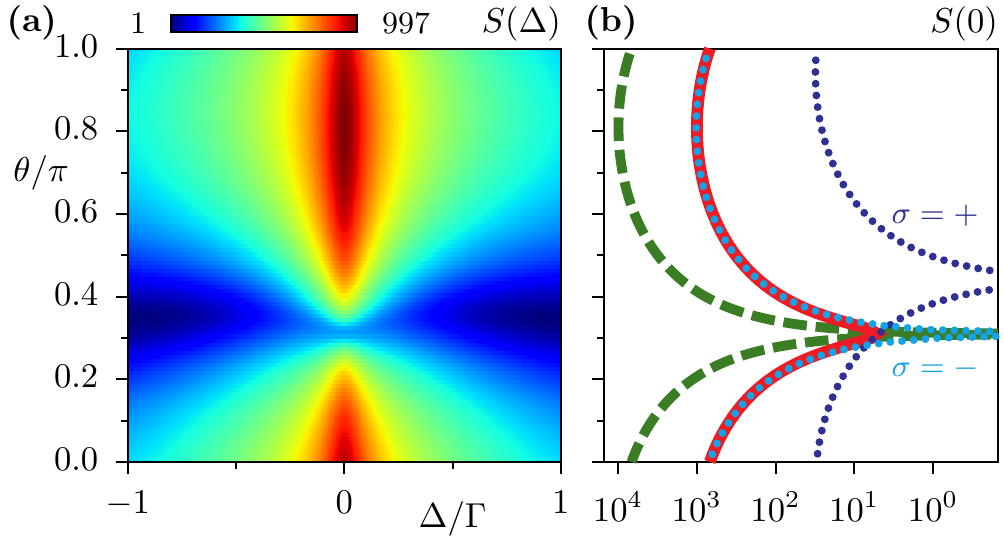}
\caption{
Output noise in presence of strong on-resonance input $B_1$, mixed with two local oscillators, at frequencies $\omega_1$ and $\omega_2$:
(a) Noise squeezing spectrum $S^\theta(\Delta)$ from sum over components in \Eq{SUV}. 
(b) Noise $S^\theta(0)$ vs. $\theta$  (red solid line)
and classical quadrature response $\overline{P_0}$, \Eq{P0} (green dashed line), 
blue dashed and light blue dotted lines indicates the supermode contributions.
Parameters are as in Fig.~\ref{fig:S11}.
}
\label{fig:Stot}
\end{figure}

A more accurate numerical computation of the nonlinear gain and squeezing spectral density supports our conclusion based on analytics about coinciding directions of the signal and total noise squeezing, although numerical value for this direction  deviates from the analytical one, $\theta \approx 0.36 \pi$, as illustrated in Fig.~\ref{fig:Stot}.

%

\subsection{Squeezed vacuum}
In this section we discuss the properties of the squeezed vacuum noise under non-degenerate parametric resonance.

In the case of two-mode squeezing, the Bogoliubov transformation, \Eq{CB}, can be written through the squeezing operator similar to the degenerate resonance \cite{SqueezingBook},
\begin{eqnarray}
c_n(\Delta) &=& e^{i\eta_n(\Delta)} S b_n(\Delta) S^\dag, \\
S &=& \exp\left[ \int_{-\infty}^\infty d\Delta' \xi(\Delta') b_1^\dag (\Delta') b_2^\dag (-\Delta') - \text{h.c.}\right] ,\nonumber
\end{eqnarray}
where $\xi(\Delta)= r(\Delta) e^{i\rho(\Delta)}$, and $\rho =  {\rm arg}\,( v/u) + \pi$, and $\eta_n = {\rm arg} \,u_n$ 
($\rho(\Delta)$ and $r(\Delta)$ are identical for both modes by virtue of the second property in \Eq{uvproperty}).
Then the squeezed vacuum wave function has the form, 
\begin{eqnarray}
|\Psi\rangle = S  |0\rangle, 
 \end{eqnarray}
and can be written explicitly using the decomposition formula \cite{Perelomov1977,Collet1988} and skipping the phase prefactor,
\begin{eqnarray}\label{}
&& |\Psi \rangle = 
 \prod_\Delta \sum_{n=0}^{\infty} {g^n(\Delta) \over \cosh r(\Delta)}
 \, | n(1,\Delta)\rangle\, | n(2, -\Delta) \rangle\,,
\end{eqnarray}
here $g(\Delta) = \tanh r(\Delta) e^{i \rho(\Delta)}$, and $|n(j,\Delta)\rangle$ is the n-photon state at the frequency 
$\omega_j + \delta$ with detuning $\Delta$. In other words, squeezed vacuum consists of an uncorrelated set of states, each being formed by correlated photon pairs of conjugated modes, 
$(1,\Delta),\; (2, -\Delta)$. 

For the four-mode squeezing, we start with the Bogoliubov transformation in the supermode basis, and present equations (\ref{cbsigma}) on the form,
\begin{eqnarray}
&& c_\sigma(\Delta) = S  b_\sigma(\Delta) S^\dag \,, \\
&& S = \exp\left[ \sum_\sigma \int_0^\infty d\Delta' \xi_\sigma(\Delta') b^\dag_\sigma(\Delta')  
b^\dag_\sigma(-\Delta')  - {\rm h.c.} \right] \,, \nonumber
\end{eqnarray}
where $\xi_\sigma(\Delta)= r_\sigma(\Delta)e^{i\rho_\sigma(\Delta)}$, $\rho_\sigma = {\rm arg}(\,v_\sigma / u _\sigma) + \pi$, and where we have omitted a phase prefactor.
Since the operator $S$ is a scalar in the mode space, the Bogoliubov transformation for the original modes  can be written through the same operator,
\begin{eqnarray}
c_n(\Delta) = S b_n(\Delta) S^\dag\,,
\end{eqnarray}
and the exponent of $S$ expressed in the original mode basis reads,
\begin{eqnarray}
\sum_{\sigma, kl} \int_0^\infty d\Delta' \xi_\sigma(\Delta') \mathfrak{U}^T_{\sigma k} \mathfrak{ U}^T_{\sigma l} b^\dag_k(\Delta')  b^\dag_l(-\Delta') - \text{h.c.}\,. \nonumber
\end{eqnarray}
 The four-mode squeezed vacuum wave function has the form,
\begin{eqnarray}\label{Psiabove}
&& |\Psi\rangle  =   \prod_{\Delta>0} {1 \over \cosh r_+(\Delta)\cosh r_-(\Delta)} \nonumber\\
&& \times  \exp\left[{g_+ + g_- \over2} \left( e^{i\psi}b^\dag_1(\Delta)b^\dag_1(-\Delta)  
+ e^{-i\psi}b^\dag_2(\Delta)b^\dag_2(-\Delta)\right) \right] \nonumber\\
&&\times \exp\left[{g_+ - g_- \over2} \left(b^\dag_1(\Delta)b^\dag_2(-\Delta) + b^\dag_2(\Delta)b^\dag_1(-\Delta)\right)\right] |0\rangle\,, \nonumber
\end{eqnarray}
with $g_\sigma = \tanh r_\sigma  e^{i \rho_\sigma}$.
This four-mode vacuum consists of a set of independent correlated photon pairs that belong to all possible combinations of the state quartet, $(1,\pm\Delta)$, $(2,\pm\Delta)$. 
It is worth noting that the admixture of the pairs from the same  mode (second line in the equation) is controlled by the intracavity field, $A_n$, and sensitive to the phase of the strong field $\psi$,  while the admixture of the pairs that belong to the different modes (third line) is predominantly controlled by the flux pumping, $\epsilon$.  

\section{Frequency conversion}
\label{Conversion}

In this last section we will consider the frequency conversion regime.
In this regime the pump frequency is chosen close to difference of the cavity resonances,
$\Omega = \omega_2 - \omega_1 + 2\delta$, $\omega_2>\omega_1$, and the dynamics of the system is described by  \Eq{eq_upc:EOM_A1_hyb} in the rotating frame, $\omega_1-\delta$ and $\omega_2+\delta$.  

Consider a nonlinear response to an input consisting of equally detuned signals in each mode, 
$B_n(t) = B_n(\Delta) e^{-i\Delta t}$.
The classical dynamical equations for the intracavity field amplitudes, $A_n(t) = A_n(\Delta)  e^{- i\Delta t}$, read,
\begin{eqnarray}\label{eq:EOMconversion}
&&\!\!\!\!\!(\Delta + \zeta_1 + i \Gamma_1) A_1(\Delta)  + \epsilon A_2(\Delta)  = \sqrt{2\Gamma_{10}} B_1(\Delta)  
\nonumber\\
&& \!\!\!\!\!(\Delta + \zeta_2 + i \Gamma_2) A_2(\Delta)  + \epsilon A_1(\Delta)  = \sqrt{2\Gamma_{20}} B_2(\Delta) .
\end{eqnarray}
Here $\zeta_n$ are defined slightly differently compared to the amplification case,
\begin{eqnarray}\label{eq:zetas_conversion}
\zeta_1 = - \delta + \alpha_1 |A_1|^2 + 2\alpha |A_2|^2 \nonumber\\
\zeta_2 =   \delta+ \alpha_2 |A_2|^2 + 2\alpha |A_1|^2\,.
\end{eqnarray}
Solving for $A_n$  we  derive the input-output relation,
\begin{equation}\label{Vconversion}
\left(
\begin{array}{c}
  C_1      \\
  C_2     
\end{array}
\right)
=  {\cal V}(\Delta)
\left(
\begin{array}{c}
  B_1      \\
  B_2     
\end{array}
\right)\,,
\end{equation}
where ${\cal V}(\Delta)$ is the intermode scattering matrix with matrix elements,
\begin{eqnarray}\label{Vnm_conversion}
{\cal V}_{11}(\Delta) &=& 1 - {2i\Gamma_{10} ( \Delta+i\Gamma_2 +\zeta_2) \over D(\Delta) }  \nonumber \\
{\cal V}_{22} (\Delta)&=& 1 - {2i \Gamma_{20} ( \Delta+i\Gamma_1 +\zeta_1) \over { D}(\Delta) }   \\
{\cal V}_{12} (\Delta)&=&  {2 i \epsilon\sqrt{\Gamma_{10}\Gamma_{20}}\over {D}(\Delta) } =  {\cal V}_{21} (\Delta)   \nonumber\\
D(\Delta) &=& (\Delta+i\Gamma_1+\zeta_1)(\Delta+ i\Gamma_2+\zeta_2) - \epsilon^2\,. \nonumber
\end{eqnarray}
%
%
\begin{figure}[tb]
\includegraphics[width=\columnwidth]{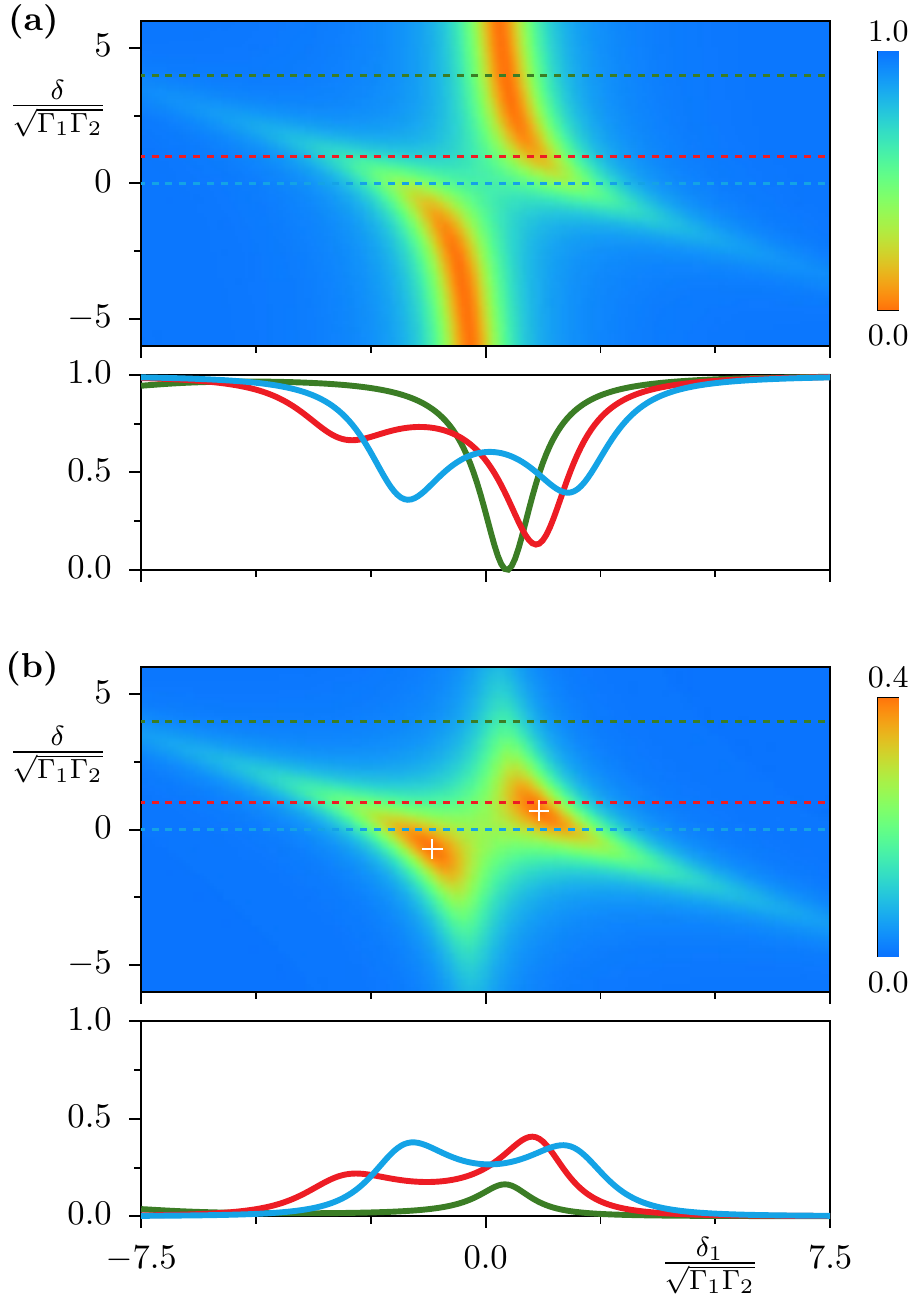}%
\caption{
Linear frequency conversion as function of input signal detuning $\delta_1$
and pump detuning $\delta$. (a) Reflection coefficient $|{\cal V}_{11}(\delta_1)|^2$ quantifies response in the input mode, it exhibits an avoided crossing of a loss resonance; solid color lines on lower panel correspond to respective cuts at different $\delta$ on upper panel  indicated with dashed lines. (b) Conversion coefficient  $|{\cal V}_{12}(\delta_1)|^2$ quantifies response in the second mode emerging at detuning, $\delta_2 = 2\delta+\delta_1$; 
white markers indicate points of maximum conversion. Observe different color codes on main panels in  (a) and (b). 
($\Gamma_{20}=3\Gamma_{10}$, 
$\Gamma_{1} = 1.8\Gamma_{10}$,
$\Gamma_{2} = 4\Gamma_{10} = 4/3 \Gamma_{20}$,
$\alpha_2=3\alpha_1$,
$\alpha_1 = \sqrt{\Gamma_{10}\Gamma_{20}}/100$).
}
\label{fig:lineargain_dw1_dpD_internalloss__frconv}
\end{figure}


In the absence of internal damping, $\Gamma_n = \Gamma_{n0}$, this matrix is unitary,
\begin{equation}\label{conservation}
{\cal V}{\cal V}^\dag = {\cal V}^\dag{\cal V} =1, \quad |{\cal V}_{22}| = |{\cal V}_{11}|, \quad |{\cal V}_{12}| = |{\cal V}_{21}|\, ,
\end{equation}
which ensures preservation of the photon number during the conversion. 

We note that the unitary property of the scattering matrix is to be considered with care since the transformation, \Eqs{Vconversion}--(\ref{Vnm_conversion}), is not a linear operation. It implicitly depends on the input via the intracavity  Kerr effect, \Eq{eq:zetas_conversion}. Thus this transformation cannot be automatically extended to the quantum regime.

\begin{figure}[tb]
\includegraphics[width=\columnwidth]{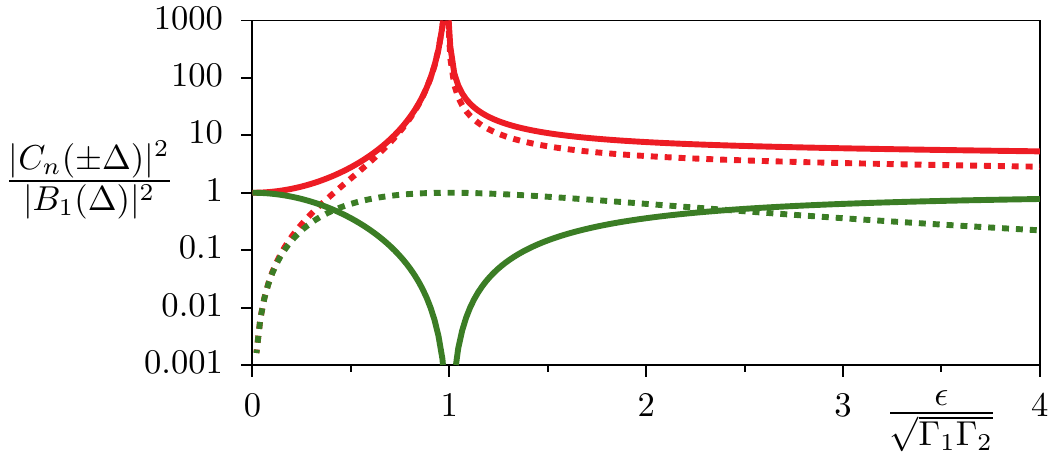}%
\caption{
Efficiency of the (linear) parametric conversion vs pump strength $\epsilon$ (green lines), characterized with conversion (dashed), and reflection (solid) coefficients. This is compared with the gains in the parametric amplification (red lines) of a signal (solid), and idler (dashed).   
($\delta=0$, $\delta_1=0$,
$\Gamma_{2}=3\Gamma_{1}$, 
$\alpha_2=3\alpha_1 = 3\sqrt{\Gamma_{1}\Gamma_{2}}/100$).}
\label{fig:lineargain_frconv_paramp}
\end{figure}

The denominator in \Eq{Vnm_conversion} never turns to zero, thus no intrinsic instability occurs, and a sufficiently small input induces a small intracavity field. In this case a linear approximation is appropriate, $\zeta_{1,2} \approx \mp \delta$, 
and  \Eqs{Vconversion}--(\ref{Vnm_conversion}) apply to the quantum regime.

In Fig.~\ref{fig:lineargain_dw1_dpD_internalloss__frconv}
we show the linear reflection (a), and conversion (b) spectra,  
$|{\cal V}_{11}(\delta_1)|^2$, and $|{\cal V}_{12}(\delta_1)|^2$,
of the parametric conversion process for the input signal, $B_1(\delta_1)$, versus input and pump detunings. 

For pump detuned far away from the resonance, $|\delta| \gg \Gmean$, the spectrum is dominated by a loss resonance centred at $\delta_1 = 0$ (green dashed line and green curve). Close to the parametric resonance, 
$\delta < \Gmean$, the intermode coupling appears as an avoided crossing  (red and blue dashed lines and curves on panel (a)) that is accompanied 
by the emergence of the converted signal, panel (b). The points of maximum conversion are indicated with white markers.

A full reciprocal conversion between the modes is possible in the absence of internal losses. The criterion is given by the zero reflection coefficient,  $|{\cal V}_{11}|^2=0$. The corresponding conditions read, for the linear regime, 
\begin{eqnarray}\label{full_conversion}
\epsilon^2 = \Gamma_1\Gamma_2 \left( 1+ {4\delta^2\over (\Gamma_2 - \Gamma_1)^2}\right), \quad
\Delta =  {\Gamma_1 + \Gamma_2 \over \Gamma_2- \Gamma_1}\,\delta\,. \nonumber
\end{eqnarray}

It is instructive to compare these equations to the ones for the parametric instability in the amplification regime, \Eqs{eq:epsilon_thresh}--(\ref{eq:Delta_thresh}): both criteria coincide at the zero pump detuning, $\delta=0$.  At finite pump detuning full conversion is still possible, but in this case the input must be detuned accordingly. The efficiency of the frequency conversion at different pump strengths is illustrated in Fig.~\ref{fig:lineargain_frconv_paramp}.


\section{Conclusion}
In conclusion, we have studied non-degenerate parametric resonance in a tunable superconducting microwave cavity. 
The main focus was put on the nonlinear properties of the resonance, stemming from the nonlinear current-phase dependence of the SQUID controlling the cavity. We analyzed nonlinear gains in the strong amplification regime at the parametric oscillation threshold, and evaluated the maximum values of the gains. We showed that the linear response of  empty cavity has the property of two-mode squeezing, while the response of the cavity filled with radiation has a four-mode structure. We identified the parametric oscillation regime and showed that  the oscillation frequencies deviate from the cavity resonances, the deviations growing with the strength of the pump. A continuous degeneracy of the oscillator state with respect to the oscillation phase causes divergence of the linear response at the oscillation frequencies for all pump strengths above the threshold. We found that injection of a weak on-resonance signal locks the oscillation phase and makes the response regular. 
We also calculated noise squeezing spectral densities in the two-mode and four-mode regimes, and found that the output signal to noise ratio in the  four-mode regime can significantly exceed  the input value.
Finally, we investigated the parametric frequency conversion and identified the conditions for full and reversible conversion.

\subsection*{ Acknowledgement}
The authors are grateful to Andreas Bengtsson, Per Delsing, and  Giulia Ferrini for useful discussions. Support from Knut and Alice Wallenberg Foundation is gratefully acknowledged.

\begin{appendix}{}

\section{Parametric oscillation state}
\label{sec:limitcycle_downconversion}

To find solutions of \Eq{eq:EOM_class_ampl} for a self-sustained oscillation above the parametric threshold, we consider the ansatz,
$A_{1,2}(t) = r_{1,2} e^{i\theta_{1,2}}e^{\mp i\Delta_0 t}$, and substitute it into the homogeneous equation,
\begin{eqnarray}
(\Delta_0 + \zeta_1 + i\Gamma_1 ) r_1 + \epsilon r_2 e^{-i\Theta} =0 \nonumber\\
(-\Delta_0 +\zeta_2 + i\Gamma_2 ) r_2 + \epsilon r_1 e^{-i\Theta} =0\,, 
\end{eqnarray}
where $\Theta = \theta_1+ \theta_2$. Then we separate real and imaginary parts of the equations.

The imaginary parts read,
\begin{eqnarray}\label{Im}
&&\Gamma_1  r_1 - \epsilon r_2 \sin\Theta =0  \nonumber\\
&&\Gamma_2  r_2 - \epsilon r_1 \sin\Theta =0\,,
\end{eqnarray}
and yield,
\begin{eqnarray}\label{sin}
{r_1\over r_2} = \sqrt{\Gamma_2\over \Gamma_1}, \quad  
 \sin\Theta = {\sqrt{\Gamma_1\Gamma_2}\over \epsilon} > 0 \,.
\end{eqnarray}
The real parts read,
\begin{eqnarray}\label{Re}
(\Delta_0 + \zeta_1 ) r_1 + \epsilon r_2 \cos\Theta &=&0  \nonumber\\
(-\Delta_0 + \zeta_2)  r_2 + \epsilon r_1 \cos\Theta &=& 0 \,,
\end{eqnarray}
from which we extract the relation,
\begin{eqnarray}
{\Delta_0 + \zeta_1\over -\Delta_0 + \zeta_2}= {r_2^2\over r_1^2} = {\Gamma_1\over \Gamma_2} \,,
 \end{eqnarray}
that defines the detuning $\Delta_0$,
\begin{eqnarray}
\Delta_0 = {\zeta_2\Gamma_1 - \zeta_1\Gamma_2 \over \Gamma_1 + \Gamma_2} \,.
\end{eqnarray}
Using this equation we compute,
\begin{eqnarray}
&& \Delta_0 + \zeta_1 = {(\zeta_1+\zeta_2) \Gamma_1\over \Gamma_1 + \Gamma_2} \,,
\end{eqnarray}
and substituting in  \Eq{Re} we derive an equation for $\cos\Theta$,
\begin{eqnarray}\label{cos}
\cos\Theta = 
- \,{\zeta_1+\zeta_2 \over \epsilon}{\sqrt{\Gamma_1 \Gamma_2} \over \Gamma_1 + \Gamma_2} 
= \pm {\sqrt{\epsilon - \Gamma_1\Gamma_2}\over \epsilon}  \,. 
\end{eqnarray}
Writing explicitly $\zeta_n$ through $r_n$ in this equation and excluding $r_2$ using \Eq{sin}, we finally get,
\begin{eqnarray}\label{r1}
r_1^2 = {2(-\delta \mp \delta_{\rm th}) \Gamma_2\over \alpha_1\Gamma_2 + \alpha_1\Gamma_2 
+ 2\alpha(\Gamma_1 + \Gamma_2)}\,,
\end{eqnarray}
where we also used \Eq{delta_thresh}. The upper (lower) sign in \Eqs{cos} and (\ref{r1}) corresponds to the unstable (stable) solution.

\end{appendix}


\end{document}